\documentclass[twocolumn,showpacs,preprintnumbers,amsmath,amssymb,superscriptaddress]{revtex4}
\usepackage{graphicx}% Include figure files
\usepackage{dcolumn}% Align table columns on decimal point
\usepackage{bm}% bold math
\usepackage{longtable}% bold math
\usepackage{amsmath}
\usepackage{amssymb}
\usepackage{mathtools}
\usepackage{booktabs}
\usepackage{tikz}
\def\quad{\hskip1em\relax}
\def\qquad{\hskip2em\relax}

\usepackage{enumerate}

\newcommand{\be}{\begin{equation}}
\newcommand{\ee}{\end{equation}}
\newcommand{\ba}{\begin{eqnarray}}
\newcommand{\ea}{\end{eqnarray}}

\usepackage{graphicx}

\def\Be'{\beta_\mu^{'}}

\def\<{\bigl\langle}
\def\>{\bigr\rangle}

\begin{document}

%\preprint{APS/123-QED}

\title{Insights in Economical Complexity in Spain: \\ the hidden boost of migrants in international tradings.}

\author{Elena Agliari}
\affiliation{Dipartimento di Fisica, Sapienza Universit\`{a} di Roma, Italy}
\author{Adriano Barra}
\affiliation{Dipartimento di Fisica, Sapienza Universit\`{a} di Roma,  Italy}
\author{Andrea Galluzzi}
\affiliation{Dipartimento di Matematica, Sapienza Universit\`{a} di Roma, Italy}
\author{Francisco Requena-Silvente}
\affiliation{Department of Economics, Sheffield University, United Kingdom}
\author{Daniele Tantari}
\affiliation{Dipartimento di Matematica, Sapienza Universit\`{a} di Roma, Italy}

\date{\today}

\begin{abstract}
We consider extensive data on Spanish international trades and population composition and, through statistical-mechanics and graph-theory driven analysis, we unveil that the social network made of native and foreign-born individuals plays a role in the evolution and in the diversification of trades.
Indeed, migrants naturally provide key information on policies and needs in their native countries, hence allowing firm's holders to leverage transactional costs of exports and duties. As a consequence, international trading is affordable for a larger basin of firms and thus results in an increased number of transactions, which, in turn, implies a larger diversification of international traded products. These results corroborate the novel scenario depicted by ``Economical Complexity", where the pattern of production and trade of more developed countries is highly diversified. We also address a central question in Economics, concerning the existence of a critical threshold for migrants (within a given territorial district) over which they effectively contribute to boost
international trades: in our physically-driven picture, this phenomenon corresponds to the emergence of a phase transition and, tackling the problem from this perspective, results in a novel successful quantitative route.
Finally, we can infer that the pattern of interaction between native and foreign-born population exhibits small-world features as small diameter, large clustering, and weak ties working as optimal cut-edge, in complete agreement with findings in ``Social Complexity''.
\end{abstract}

%Keywords: statistical mechanics, spin-glasses, complex networks, theoretical immunology.

%\pacs{05.40.Fb, 02.50.Cw, 02.50.Ey}

\pacs{89.65.Ef, 89.65.-s, 05.40.-a, 05.70.Fh} \maketitle

\section{Introduction}\label{intro}

In this work we aim to merge recent findings in Social Complexity \cite{Barra-ScRep,Agliari-NJP}  with those achieved in  Economical Complexity \cite{barabasi2,barabasi3},
%exploiting and formalizing the ides exposed in \cite{francisco1,francisco2}
in order to deepen our understanding of socio-economical behaviors observed in developed societies. In particular, we examine the role of migratory fluxes on the economical diversification and international trading of  the hosting  countries.
%In particular we will show that key features of social dynamics regarding local, microscopic, interactions among binary decision makers (i.e. Ising spins), discovered by pioneers as Granovetter Ä}, Watts and Strogatz \{}, and  Brock and Daurlauf \cite{}, are perfectly in agreement with (and indeed act as a social fingerprint of) the resulting Economical Complexity highlighted by Hidalgo, Klinger and Barabasi \cite{}, and some of the present authors \cite{},  when testing the diversification of products a society imports or exports.

The emergence and the fitness of economical diversification is nowadays still questioned: classical economic theories prescribe specialization of industrial production for more performing countries \cite{old1,old2}, while recent studies \cite{barabasi2,barabasi3,pietronero1,pietronero2} show that diversification of products plays a key role in modern economies. Quoting Hidalgo, Klinger, Barabasi and Hausmann {\em ``inspection of the country databases of exported products shows that successful countries are extremely diversified, in analogy with biosystems evolving in a competitive dynamical environment"} \cite{barabasi2}. Oversimplifying, the key idea to explain such a diversification is that, if the factors (e.g., technology, capital, institutions, skills) necessary  for a country to produce a good are (partially) shared with another good,  it will be likely that both goods will be produced \cite{barabasi2}.
\newline
Here, we address a closely related problem: we investigate the diversification of the production of a country by looking at its exports and connecting  {\em diversification in trades} with {\em social complexity} beyond {\em economical complexity}. In particular, we quantitatively show that stocks of foreign migrants play a crucial role in the establishment of international trades of diversified products, thus contributing to explain the genesis of the Hidalgo, Klinger, Barabasi and Hausmann picture. In a nutshell, our results (in agreement with recent literature \cite{egger,head,hunt,ozgen,partridge,rashidi} \footnote{Exception may arise due to peculiar historical and/or colonial traditions \cite{girma,gould}}), suggest that social interactions between native and foreign-born populations allow transferring to local firms a crucial knowledge about policies, needs and duties existing in the foreign countries. Remarkably, this information, coupled with firms' holder capabilities, permits to decrease the overall potential costs of trading thus allowing a larger number of firms to appear in the global market, which, in turn, implies broader and diversified trades.

Thus, our claim is that the interaction network between migrants and natives spreads the \emph{social capital} (i.e. the collective resources of the community, including information, expertise and skills) and this enhances the extensive margin of trades, which, in turn, acts as a boost in the diversification of the exported products.

In order to prove these statements, we introduce a statistical-mechanics scaffold (where data can be rationally framed)  and, step by step, we check for the empirical confirmation of our assumptions and our theoretical results, by analyzing the test case of Spain. In fact, this country has experienced a (well-documented) influx of migrants since $1998$ with a very rapid increase during the period $2000-2008$ \cite{francisco1,francisco2,francisco3,Barra-ScRep,Agliari-NJP} and this constitutes an ideal context to investigate the role of immigrants in creating new trade relationships.
\newline
More precisely, our work is structured as follows.
\newline
In the first part, devoted to the statistical mechanical analysis, we introduce the simplest possible model (i.e., a minimal Hamiltonian) that relates two parties: foreign-born and native people living in a given district of the country. As a result of the interaction between the two parties, natives will -stochastically- decide whether to trade with the country of origin of immigrants. Remarkably, we prove that this model belongs to the class of copying-model \cite{Ghirlanda1,Ghirlanda2}, or single-party ferromagnets in the jargon of statistical physics, where native decision-makers alone come to play and they spontaneously behave in an imitative way.  Through this approach we are able to quantify the role of immigration in the volume of trades and to include this phenomenon in the framework of the phase transitions. Within this setting, we can also test empirically whether a critical mass of migrants in needed in order to ensure that a positive pro-trade effect of migration exists \cite{francisco3,peri}, in agreement with the pioneering suggestions by Gould \cite{gould} and, more recently, with the non-linear theories driven by Chaney's distorted gravity scheme \cite{caney}.  Our theoretical findings  predict  a non-linear dependence, encoded by an hyperbolic tangent, for exports to a given foreign country versus the percentage of immigrants hailing from that country, and are successfully checked by comparison with the Spanish dataset. We conclude the first part of the paper by proving the existence of a net and robust  correlation between the degree of product-destination diversification of exports (measured in terms of the Herfindhal index) and the number of migrants as a fraction of the total population.
%\newline
%Finally we move to analyze the last point, namely the existence of a threshold in migrant's density within the hosting provinces before their presence starts to boost international trading. However our results on this point are not conclusive: we find that for roughly half of the analyzed test cases (for both exporting provinces and importing countries) indeed the existence of a critical mass is confirmed (and remarkably it appears to be practically universal, i.e. independent both by the exporting province and the importing country), while for the remaining cases this threshold is absent.

Finally, our theory also allows us to infer the topological structure of the host society, and this is addressed in the second part of the paper.
Interestingly, we find that the society displays small-world features and recovers the Granovetter theory of weak ties \cite{Granovetter-1973,Granovetter-1983,Barra-PhysA2012}. Incidentally, we notice that this is also compatible with recent researches investigating the role of immigrant integration in labor markets \cite{Damm-2013}.
%In fact, as shown in \cite{Damm-2013}, unemployed people with many employed acquaintances have a higher job finding rate.

\section{Results}

Before introducing our model, a few points must be clarified (and empirically proven to hold):
\begin{itemize}

\item Our theory, developed within a classical statistical mechanical perspective, is set at a microscopic level and it accounts for an ensemble of native ``decision makers'', whose behavior (i.e., the propensity to undertake an international trade) can be affected by the interaction with migrants. However, the theoretical outcomes of such a model are  compared with available data on international trades performed by firms: in principle, it is not obvious that we can switch from the microscopic level (i.e. decision makers), where the whole theory lies, to the {\em mesoscopic level} (i.e. firms), where the data analysis is performed. This is allowed if and only if there exists a linear proportionality between the total population  and the total amount of firms. Luckily, this is the case in Spain for the considered time window (1998-2012), as corroborated by empirical findings shown in Fig.~\ref{fig:FirmPop}. Thus, as far as scalings are concerned, we can exploit the theoretical predictions for the average behavior of decision makers (stemming from the statistical-mechanics model) to describe the expected attitude of firms (that we infer from empirical data).
%Thus, as far as scaling and proportionality laws must be inferred, our procedure is allowed.

\begin{figure}[h!]
\includegraphics[width=0.5\textwidth]{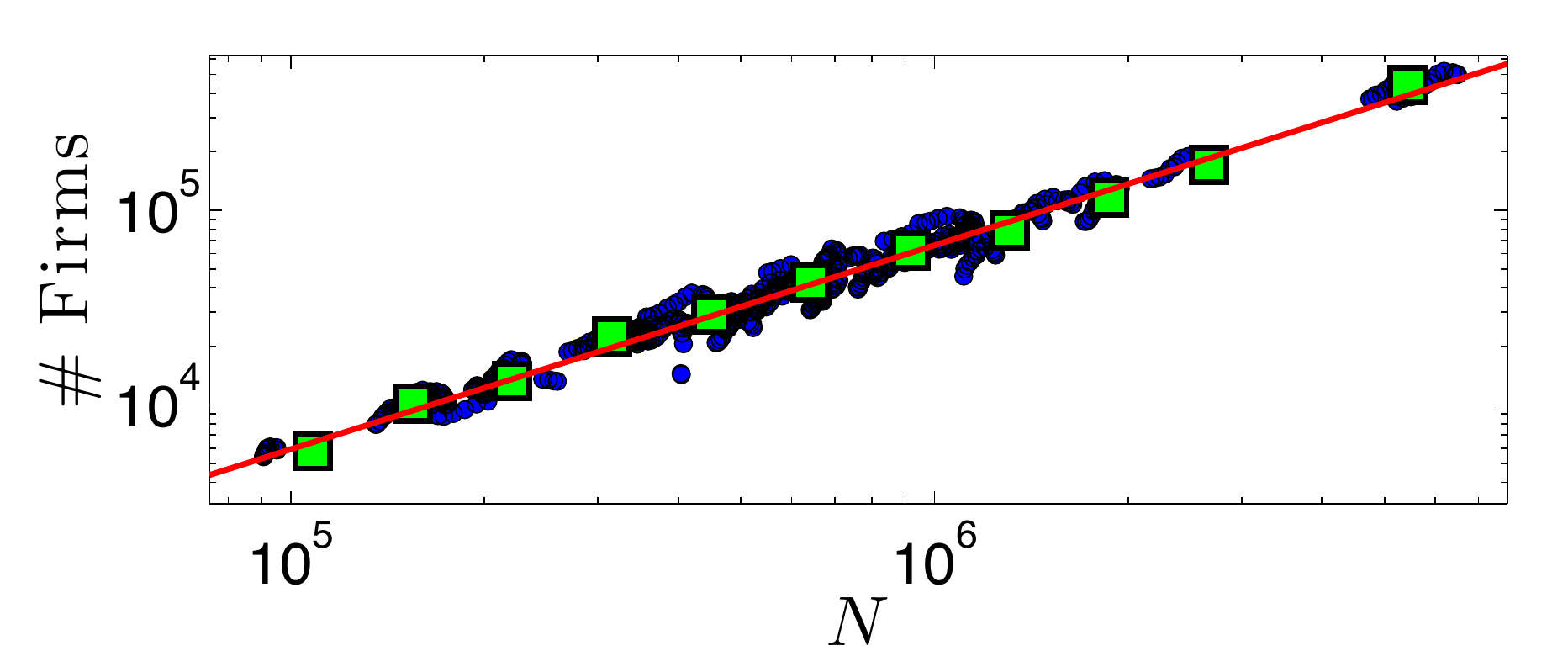}
\caption{Each data point (blue bullet) represents the number of firms versus the population of a given Spanish province (out of $50$) for a given year (in the interval $1998-2012$). The linear proportionality of these quantities is highlighted by binned data (green squares), whose best fit is given by a linear law (red solid line) with slope $\approx 1.02 \pm 0.03$ and goodness $R^2 \sim 0.99$.}
\label{fig:FirmPop}
\end{figure}

\item The total amount of trades $Y$ is usually defined in terms of two contributions: the amount of firms that perform international trading (i.e. extensive margin $Y_{ext}$) and the amount of money each firm moves in any transaction (i.e. intensive margin $Y_{int}$), namely $Y=Y_{ext}\cdot Y_{int}$, or, in a logarithmic scale, $\log Y = \log Y_{ext} + \log Y_{int}$. Chaney has shown that a reduction in fixed trade costs has a positive impact on $Y_{ext}$ \cite{caney}; Peri and Requena have shown that migrants have a positive effect on the extensive margin of trade in Spain, hence deriving that migrants facilitate trade mainly by reducing the fixed costs of exporting \cite{francisco1,francisco3}. On the other hand, the intensive margin of trades seems to be poorly affected by migration stocks. Thus, our theory is actually devoted to  capture  the evolution of $Y_{ext}$.

 %and logical sequence (linear proportionaly) we will use is that the order parameter from statistical mechanics (called $m$ later on) is proportional to the extensive margin $Y_{ext}$, that in turn is proportional to the amount of exports $E$. Thus we will compare scalings and proportionalities of $m$ versus scalings and proportionalities of $E$.

%\item \emph{When looking at $Y_{ext}$, we are implicitly accounting for relatively small firms (C'ERA UN FILE COL NUMERO DI PERSONE NELLE FIRM). For such structures, the extent of trades (in terms of the money involved) exhibits a characteristic scale $M$. In general this is not the case and we have power-law behaviours. We can therefore establish a further correspondence between the overall number of trading firms and the extent of trades $Trades$, as $Trades \propto M Y_{ext}$. PROVE A SOSTEGNO DI QUESTA TESI... We can therefore exploit the the theoretical predictions for the average behavior of decision makers (stemming from the statistical-mechanics model) to describe the expected attitude of the volume of trades (which correspond to the available data).}
%

\item The database available reports about the total volume of transaction, that is $\log Y$.
%In order to infer about a possible role of migrants in boosting international trading between the hosting country and the country of origin, by constraints on data, we will be forced to deal with the total volume of transaction, that is $\ln Y$.
As a consequence, we first need to prove that the expected linear proportionality between $\log Y$ and $\log Y_{ext}$ is fulfilled by our data, such that, later, we will be authorized to analyze the evolution of $\log Y$ as a function of migrant density inside the host country in order to extrapolate an analogous scaling for $\log Y_{ext}$ too. This proportionality is robustly checked as shown in Fig.~\ref{fig:TLC}.

%\item
%The volume of the overall trades include both imports, $I$, and exports, $E$. However, as imports and exports exhibit a linear correlation (see Fig.~\ref{fig:ImpExp}) we will focus only on exports for the sake of simplicity.

\begin{figure}[h!]
\includegraphics[width=0.45\textwidth]{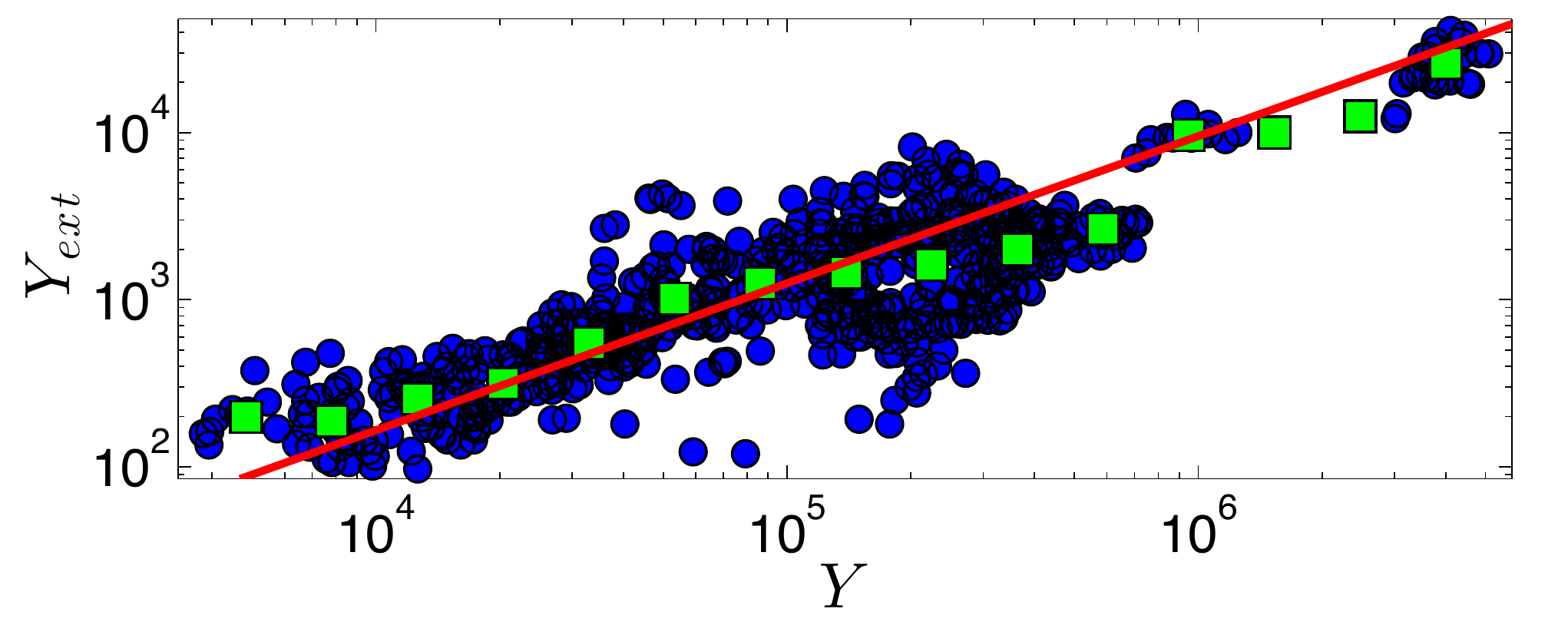}
\caption{Each data point (blue bullet) represents the number of exporting firms $Y_{ext}$ versus the overall extent of trades $Y$ for a given Spanish province (out of $50$) for a given year (in the interval $1998-2012$). Binned data (green squares) are best-fitted by a straight line (red solid line) $y = a x$, being $a \approx 0.006$, and $R^2 \approx 0.89$.}
\label{fig:TLC}
\end{figure}

%\begin{figure}[h!]
%\includegraphics[width=0.45\textwidth]{ImpExp2.pdf}
%\caption{Each data point correspond to a given Spanish province at a given year. The green solid squares have been obtained by binning the whole set of data, while the red solid line has been obtained by fitting data according to the linear law $y = p_1 x + p_2$, in such a way that we expect exports $E$ scaling with imports $I$ like $E \sim I^{p_1}$.  Data are fitted through $\log (E) =p_1 \log(I) +p_2$, with $p_1=1.00 \pm 0.02$ and $p_2=0.06 \pm 0.01$ and an $R^2\sim 0.98$.}
%\label{fig:ImpExp}
%\end{figure}
%
\end{itemize}

\subsection{PART ONE: Insights from Statistical Mechanics}

First, we need to set a proper {\em length-scale}: as the migration-trade relation is known to be an {\em in-province} phenomenon \footnote{This means that exports from a province to a given foreign country do not receive any stimuli by immigrants coming from that country but living in a different province} \cite{francisco2}, we fix the degree of resolution at the provincial level. Then, for any arbitrary province, we denote with $N$ its population and notice that the $N$ individuals can be divided into two groups: $N_1$ natives and $N_2$ foreign-born, being $N_1+N_2=N$. We also define
\be \label{eq:gamma}
\gamma \equiv \frac{N_2}{N}, \ \ 1-\gamma \equiv \frac{N_1}{N},
\ee
measuring the relative size of the two groups and we introduce $\Gamma \equiv \gamma(1-\gamma)
$ too, the latter representing the normalized number of cross links between the two communities: note that for small $\gamma$ (and this is the case for Spain), $\Gamma \sim \gamma$.
%\newline
%We keep the focus at the level of provinces because empirical findings \cite{francisco2} firmly show that the migration-trade relation is an {\em in-province} phenomenon, that is to say, exports from a province to a given country do not receive any stimuli by immigrants coming from that country but living in a different province.
%

Moreover, we introduce variables (i.e. spins), referred to as $\{\sigma_i\}_{i=1}^{N_1}$ and $\{z_{\mu}\}_{\mu=1}^{N_2}$, respectively, such that $\sigma_i\in\{-1,+1\}$ represents the propensity of the native agent $i$ to establish ($\sigma_i = +1$) or not establish ($\sigma_i = -1$) a trade, while the variables $z_{\mu}$ represent the quantity of information, either positive ($z_{\mu}>0$) or negative ($z_{\mu}<0$), that the $\mu$-th immigrant can provide (regarding trading toward his/her country of origin). Otherwise stated, the ensemble $\{ z_{\mu}\}_{\mu=1}^{N_2}$ represents the \textit{social capital} of the immigrant community and, in the absence of any additional information, in a mean-field approach, it can be thought of as a collection of Gaussian variables identically and independently distributed.

\begin{figure}[tb]
\includegraphics[width=0.5\textwidth]{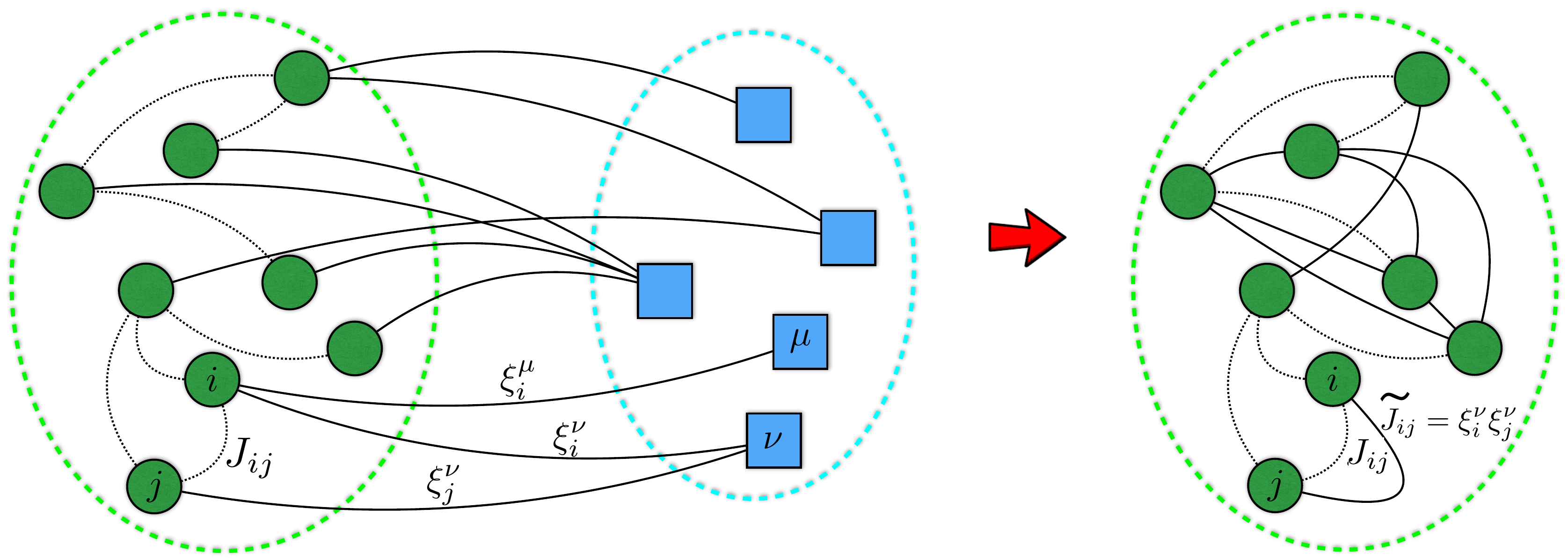}\\
\caption{Sketch of the bipartite network modeling mutual interactions between natives (left community) and immigrants (right community). The coupling between the native labeled as $i$ and the immigrant labeled as $\mu$ is denoted as $\xi_i^{\mu}$, while the coupling between two natives labeled as $i$ and $j$, respectively, is denoted as $J_{ij}$.}
\label{fig:Grafo}
\end{figure}

The diffusion of the social capital and the decisional mechanism can be now described by an Hamiltonian (i.e. a {\em cost function} in economical vocabulary) $\mathcal{H}(\sigma,z;\boldsymbol{J},\boldsymbol{\xi})$, dependent on the couplings $\boldsymbol{J}$ and $\boldsymbol{\xi}$, encoding for native-native interactions and for native-migrant interactions, respectively (see Fig.~\ref{fig:Grafo}, left panel).

Now, let us inspect in more details the interaction patterns and the resulting Hamiltonian.

The interaction between a native, say $i$, and a foreign-born, say $\mu$, is encoded by the variable  $\xi ^{\mu}_i\in \{0,1\}$ describing the presence ($\xi ^{\mu}_i=1$) or the absence ($\xi ^{\mu}_i=0$) of a connection (e.g. friend, work-mate, acquaintance, familiar) between $i$ and $\mu$.  The set of variables $\boldsymbol{\xi}$ generates the topology of the social network between immigrants and natives. Since there exist nor detailed information about individual connections, neither a broadly accepted protocol for their measure, and checking that migratory fluxes are uncorrelated (i.e. the time-scales considered are long enough and migrants comes from a wide range of countries), the most basic assumption one can then pose is simply to consider the completely general set of $\xi^{\mu}_i$ as i.i.d. aleatory variables, extracted with probability
\be \label{eq:pattern}
\mathbb{P}(\xi^{\mu}_i=1)= 1 - \mathbb{P}(\xi^{\mu}_i=0) = \frac{\xi}{N^{\theta}},
\ee
where $\theta \in (0,1)$, and $\xi \in \mathbb{R}^+$ are parameters province-dependent: in this way, properly tuning $\theta$ and $\xi$, the network recovers all the standard regimes (e.g., extreme dilution, finite connectivity, etc.) and, by fitting these parameters over the available data, we can infer the topological features of the actual Spanish network for the analyzed years \cite{Agliari-EPL2011}.
%As $\xi$ and $\theta$ are tuned we explore different dilution regimes (namely the average number of links connecting an immigrant with native community). In particular, from a fully-connected graph we can reach a sparse graph by tuning $\theta$ from $0$ to $1$.
%

Analogously, $\boldsymbol{J}$ describes the connections among natives and, at this stage, it can be assumed to be arbitrary but endowed with a well defined average value $\bar{J}$ (see also Part II for more details), which, in principle, depends on the province $p$.

Therefore, at the provincial level of resolution, the system can be described by the Hamiltonian
\be \label{eq:Hami}
\mathcal{H}(\sigma, z; \bold{J}, \bold{\xi})= -\frac{1}{N_1} \sum_{(i,j)}^{N_1} J_{ij} \sigma_i\sigma_j
-\frac {1}{ N^{1-\theta}} \sum_{i=1}^{N_1}\sum_{\mu=1}^{N_2}\xi^{\mu}_i \sigma_i z_{\mu}.
\ee
Note that in the second term in the r.h.s. of the above equation, the normalization factor $1/N^{1-\theta}$ ensures the linear extensivity of the Hamiltonian or, analogously, that the field $h_i$ acting on any spin $\sigma_i$ is $\mathcal{O}(1)$. In fact, $h_i= \frac{1}{N^{1-\theta}}\sum_{\mu}\xi^{\mu}_i z_{\mu}=\mathcal{O}(1)$, as the expected number of non-null entries in the vector $\xi_i$, namely the expected number of non-null terms in the sum, is just $\mathcal{O}(N^{1-\theta})$.\\

Before proceeding, we need to introduce a parameter $\beta$ to tune the degree of stochasticity in the system, in such a way that for $\beta \to 0$ the system behaves completely randomly, while as $\beta \to \infty$ the system deterministically relaxes to the configuration corresponding to the minimum of the cost function.
Thus, the partition function $Z$ of the model defined by the Hamiltonian (\ref{eq:Hami}) reads as
\begin{eqnarray}
\label{Z1}
Z&=&\sum_{\sigma} e^{\frac{\beta\bar{J}}{N_1} \sum_{i,j}^{N_1}\sigma_i\sigma_j} \int d\mu(z)e^{\frac{\beta}{N^{1-\theta}}\sum_{i=1}^{N_1}\sum_{\mu=1}^{N_2}\xi^{\mu}_i\sigma_iz_{\mu}}\\
\label{Z2}
&=& \sum_{\sigma}e^{\frac{\beta\bar{J}}{N_1} \sum_{i,j}^{N_1}\sigma_i\sigma_j}e^{\frac{\beta^2}{N^{2(1-\theta)}}\sum_{(i,j)}^{N_1}\sum_{\mu=1}^{N_2}\xi^{\mu}_i\xi^{\mu}_j\sigma_i\sigma_j},
\end{eqnarray}
where we called $d\mu(z)$ the standard Gaussian measure.
Crucially, by a direct comparison of the arguments in the exponents of Eqs.~\ref{Z1} and \ref{Z2}, respectively, we see that the bipartite interactions between natives and immigrants (i.e. those $\propto \sum_{i=1}^{N_1}\sum_{\mu=1}^{N_2}\xi^{\mu}_i\sigma_iz_{\mu}$ in the first line) are stored in an effective coupling $\tilde{J}_{ij}$ between couples of local decision makers alone (i.e. those $\propto \sum_{(i,j)}^{N_1}\sum_{\mu=1}^{N_2}\xi^{\mu}_i\xi^{\mu}_j\sigma_i\sigma_j$ in the second line). Such a coupling is Hebbian-like \cite{Agliari-EPL2011} as
\be \label{eq:hebb}
\tilde{J}_{ij}=\frac{\sum_{\mu=1}^{N_2}\xi^{\mu}_i\xi^{\mu}_j}{N^{2(1-\theta)}}.
\ee
Therefore, the bipartite model described in Eq.~\ref{eq:Hami} is thermodynamically equivalent to a monopartite ferromagnetic (i.e. with imitation among natives) model embedded in a random, diluted structure \cite{Agliari-EPL2011} (see Fig.~$3$, right panel). Despite the underlying graph is not fully-connected (and we will show later that, at least for the Spanish case, it is a small-world network), it is not under-percolated, hence the model still exhibits a phase transition qualitatively analogous to the one pertaining to the Curie-Weiss scenario \cite{Barra-JStat2011,bianconi}.

The ``order parameter'' for this model is given by $M(\boldsymbol{\sigma})=\frac 1 {N_1}\sum_{i=1}^{N_1} \sigma_i \mathbb{I}_{\sigma_i,1}$, namely the fraction of individuals inclined to an international trade (i.e., the amount of spins positively aligned). This order parameter is equivalent (upon  translation) to $m(\boldsymbol{\sigma})=\frac 1 {N_1}\sum_{i=1}^{N_1} \sigma_i $, namely the standard {\em magnetization} of the system (in its ferromagnetic interpretation \cite{ellis,barra0}). Now, it is worth recalling that the linear proportionality between decision makers and firms in the Spanish provinces (see Fig.~\ref{fig:FirmPop}) allows inferring only scalings and proportionality relations (but not exact values) for the amount of trading firms. Therefore, there is no loss of information in using the (mathematically more convenient) $m$ instead of $M$, and hereafter we will retain the former observable to quantify the extensive margin of trades $Y_{ext}$.
 Moreover, as explained in the previous section, the evolution in $Y_{ext}$ can be related to the evolution of trades $Y$ as a whole.

%Indeed if we introduce the set of order parameters $\{m_p^{\mu}(\sigma_p)\}_{\mu=1}^{N_2}$ (where the subscript $p$ refers to a given province $p$) as
%\be
%m_p^{\mu}(\sigma_p)=\frac{1}{C}\sum_{i=1}^{N_1}\xi^{\mu}_i\sigma_{(i_p)},
%\ee
%where $C$ normalizes with respect to the expected number of non null entries, namely
%\be
%C=N_1\mathbb{P}(\xi^{\mu}_i=1)=N_1\xi N^{-\theta}=\xi(1-\gamma)N^{1-\theta},
%\ee
By applying the standard statistical-mechanical machinery (see Appendix A for a detailed derivation), we attain the following self-consistent equation for $m$:
%\be
%m=\tanh[\beta^2\xi^2\gamma(1-\gamma)m],
%\ee
%which relates the expected amount of trading firms (and, similarly, the expected volume of international trades) with the fraction $\gamma$ of foreign-born people in the province considered.
%\newline
%Accounting also for the intra-party interaction encoded by $\mathbf{J}$ would simply imply an additional term $\beta \bar{J} m$ in the argument of the hyperbolic tangent, namely
\be\label{main-part1}
m=\tanh(\beta \bar{J}m +\beta^2\xi^2\Gamma m).
\ee
This is the main formula in this first part as, following the scaling $m \propto Y$ argued above, it relates the growth of trades with the percentage of migrants (we recall $\Gamma = \gamma(1-\gamma) \approx \gamma$). The agreement between Eq.~\ref{main-part1} and the Spanish test case is reported in Fig.~ \ref{fig:esempio} and deepened in the Data Analysis Section.

Remarkably, Eq.~\ref{main-part1} also contains information regarding the critical percentage of migrants that must be reached before they start to influence new trade relationships. To extract such information, we exploit the statistical physics know-how of {\em phase transitions}: when the argument of the hyperbolic tangent is smaller than one the only solution for Eq.~\ref{main-part1} is $m=0$. However, as the argument gets larger than one, non-zero solutions appear and we can expand the hyperbolic tangent as
\be
m\sim \beta ( \bar{J} + \beta \xi^2 \Gamma)  m -  \frac{\beta^3}{3} ( \bar{J} + \beta \xi^2 \Gamma)^3 m^3 + \mathcal{O}( m^3),
\ee
and, excluding the paramagnetic solution ($m=0$), we get
\be  \label{eq:expand}
m \sim \sqrt{\frac{3}{\beta^3( \bar{J} +\beta \xi^2 \Gamma)^3} [\beta ( \bar{J} +\beta \xi^2 \Gamma)-1]} = a  \sqrt{\Gamma-\Gamma_c},
\ee
where $a= 3 \xi^2/[\beta(\bar{J} + \beta \xi^2 \Gamma_c)^3]$ and $\Gamma_c = (1 - \beta \bar{J})/ (\beta \xi)^2$.
From the previous equation we see that as far as $\Gamma < \Gamma_c$ no real solution to this equation exists. Thus, when the percentage of migrants within a given province is smaller than $\Gamma_c$, trades can of course take place, but the related international market is not influenced by the presence of migrants within the province itself.

Three important aspects of the relation between migration and trading are thus coded in equation Eq.~\ref{main-part1}:
\begin{itemize}
\item
The relation between migrant density and growth of trades is non-linear, as these observables are related via an hyperbolic tangent.
\item
There exists a critical value for the fraction of migrants, that reads as
\be \label{eq:critical}
\Gamma_c= \frac{1 - \beta \bar{J} }{\beta^2 \xi^2},
\ee
beyond which they start to have a net effect on international trading for the host province.
Notice that $\Gamma_c$ is stochastic (via $\beta$), and, in principle, province dependent through $\bar{J}$ and $\xi$.
In fact, we stress that the previous derivation holds for any arbitrary province and, in general, the parameter set $(N, \mathbf{J},{\xi})$ is province dependent, in such a way that the outline for $m$ versus $\Gamma$ as well as the critical value $\Gamma_c$ vary with the province. However, note that, in principle, $\Gamma_c$ can be vanishing.
\item
There is a {\em saturation effect} for large enough $\Gamma$ as the hyperbolic tangent is a bounded function that eventually reaches a plateau. Exhaustion levels in bilateral exports have already been linked with migrant saturation effects as, for instance, in the experimental works discussed in \cite{egger}.
\end{itemize}

\subsubsection{Data Analysis}

We check our findings versus empirical data for the test-case of Spain. The overall dataset is obtained by merging two sources: trade data come from ADUANAS-AEAT dataset provided by Ministerio de Economia y Hacienda, and demographic data come from the Spanish Statistical Office (INE).
\newline
%If we remember that trades of the extensive margin $Y_{ext}$ (the one $\propto m$) are built by imports and exports, that we are interested in scalings only and that imports and exports are linearly related (see Fig.$2$), we can concentrate on exports only for simplicity.
We consider the time series for exports $\{ Y_{y,p} \}$ and for the fraction of immigrants $\{ \gamma_{y,p} \}$, along the range of years $y=1998,...,2012$ and for the $50$ provinces $p=1,...50$ making up the country (EUROSTAT NUTS III definition). Thus, our time range is made of $N_y = 15$ years and our geographic set is made of $N_p=50$ provinces.
\newline
Preliminarily, as we start from historical series, we check that at least one of the observables $Y$ and $\gamma$ is monotonically increasing with respect to the years $y$, and $\gamma(y)$ satisfies this request. Thus, we are allowed to invert $\gamma(y) \rightarrow y(\gamma)$ and look at the evolution of $Y$ as a function of $\gamma$, so to obtain $Y(\gamma)$ that must then be suitably binned and averaged (see \cite{Barra-ScRep} for details on this procedure).

The whole set of provinces constitutes our pool, namely we consider different provinces as independent realizations (or, otherwise stated, extractions) of the same system. This means that the trades of a given province are taken to depend only on the fraction of immigrants within the province itself. While there is general consensus on this, the consistency of such an hypothesis is shown in \cite{HerSaa}, where the authors prove that the \textit{proximity} (meant as geographical closeness) is fundamental for the diffusion of the social capital and therefore for the growth of trades.

\begin{figure}[h!]
\includegraphics[width=0.4\textwidth]{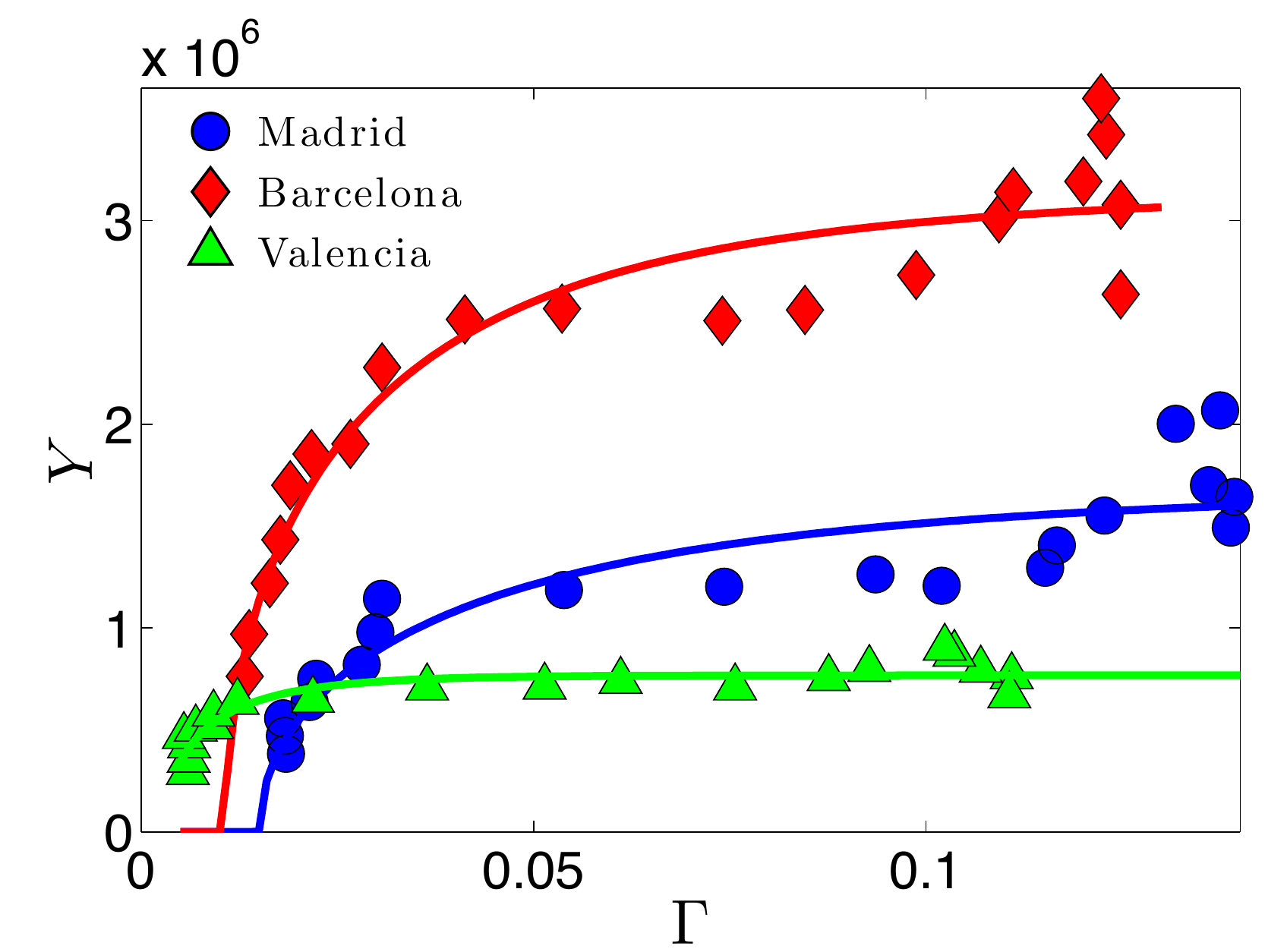}
      \caption{Exports $Y$ versus $\Gamma$ for three different provinces as explained by the legend; we choose the three largest provinces for the sake of readability and for consistency with the analysis of the following sections, however, we checked that analogous plots hold also for the other provinces. In this plot each data point corresponds to a different year. The solid lines represent the best fit according to Eq.~\ref{eq:tanh2} and the goodness of the fit is $R^2 = 0.94$ (Madrid), $R^2=0.97$ (Barcelona), and $R^2=0.95$ (Valencia).}
      \label{fig:esempio}
\end{figure}

\begin{figure}[h!]
 \includegraphics[width=0.43\textwidth]{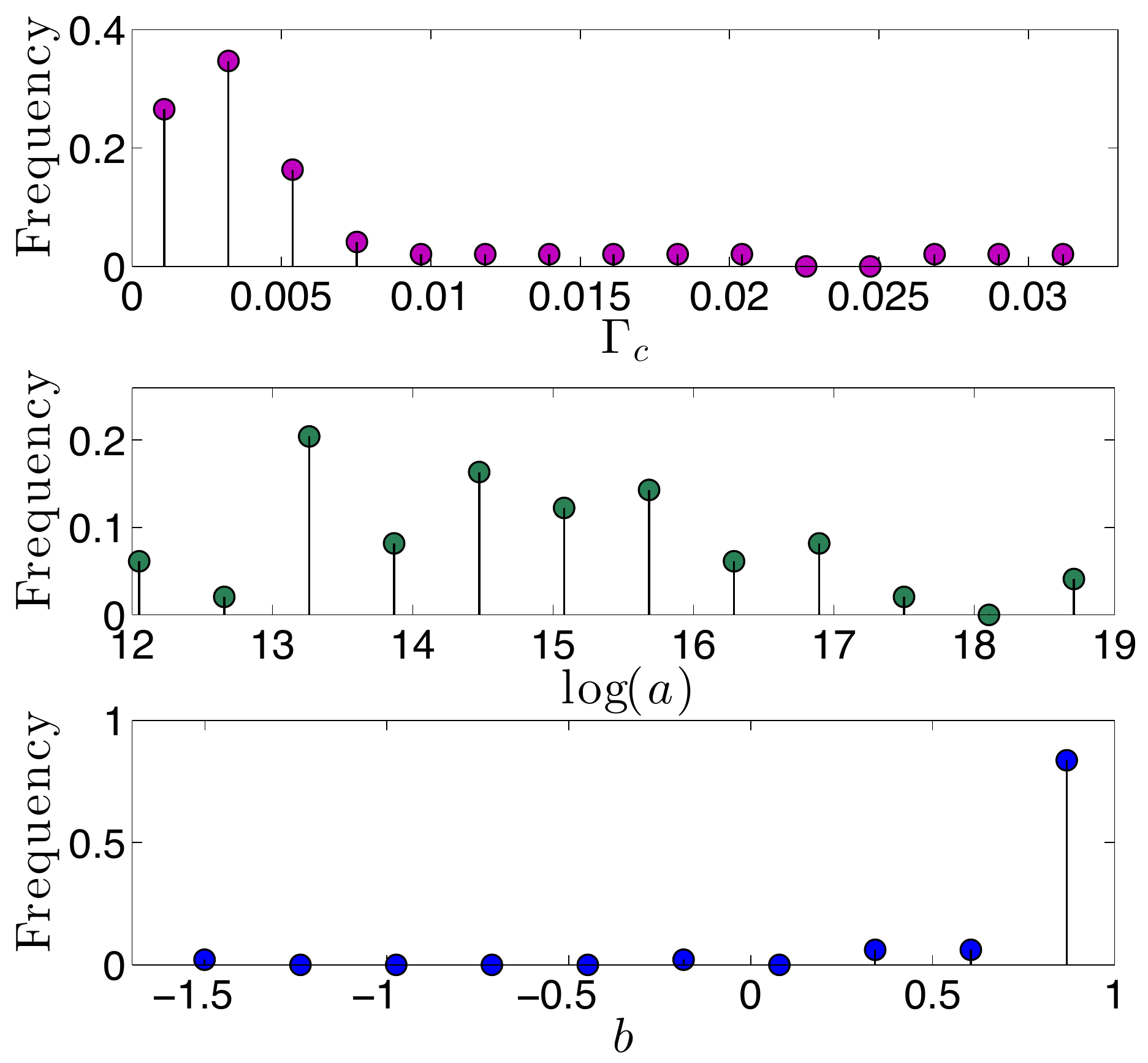}
 \caption{Histograms for the best-fit coefficients $\Gamma_c$ (upper panel), $a$ (middle panel), and $b$ (lower panel) obtained by fitting $Y_{p}$ versus $\Gamma_{p}$ according to Eq.~\ref{eq:tanh2}, for each province $p$. Notice that, due to the broad range along which  $Y_{p}$ (and, accordingly, $a$) spans, we represent the histogram of $\log(a)$.}
  \label{fig:isto}
\end{figure}

%Moreover, as highlighted by previous analysis by \cite{HerSaa,altri?}, we expect that the trades related to exports tend to increase with the fraction of immigrants. In order to highlight this point,
For each province $p$, we can measure the percentage of immigrants $\gamma_{p}$ and plot $Y_{p}$ versus $\Gamma_{p} \sim \gamma_{p}$, as shown in Fig.~\ref{fig:esempio} for some exemplary cases. Note that theoretical predictions (see Eq.~\ref{main-part1}) are in remarkable agreement with the empirical behaviour.
%
%
%\begin{figure}[h!]
%    \includegraphics[width=0.5\textwidth]{TanhRoot.pdf}
%    %\includegraphics[width=0.4\textwidth]{schiuma.eps}
%      \caption{Example of trades versus the percentage of migrants $\Gamma=\gamma (1-\gamma)$ on a linear scale:   yellow circles represent real data for the province of Cadiz, while the continuous line is the best fit achieved with the hyperbolic tangent (eq.\ref{main-part1}) with $R^2 \sim 0.98$ and the dashed line is the best fit achieved with the square root (eq. \ref{eq:expand}) with $r^2 \sim 0.94$. Note that the two curves merge quite well at the beginning of the growth (i.e., for $\Gamma < 0.005$), thus we can safety use the latter for finding the critical mass of migrants.}\label{fig:esempio2}
%\end{figure}
%
%
%%
%In Eq.~\ref{eq:radice} the parameter $a_p$ provides the characteristic scale for trades in the province $p$: given two provinces $p_1$ and $p_2$, and being $a_{p_1}>a_{p_2}$, we expect that, $E_{p_1}(\gamma) > E_{p_2}(\gamma)$, at least for sufficiently large values of $\gamma$. On the other hand, Eq.~\ref{eq:radice} becomes unfeasible as $\Gamma<b_p$ and, according to the statistical-mechanics picture, below that threshold exports are expected to be migrants-independent: $b_p$ provides a critical threshold  of the immigrant population such that  below that value their presence does not induce any significant increase in international trades by firms within their province of origin.
%
%By performing the fit over all provinces we get a set of best-fit coefficients $a_p$ and $b_p$, whose distributions are depicted in Fig.~$6$.
We performed extensive fits over all the provinces available according to Eq.~\ref{main-part1}, which we report hereafter as
\be \label{eq:tanh2}
m = \tanh \left [\left( \frac{(1-b)\Gamma}{\Gamma_c}  +b \right) m \right],
\ee
where we highlighted the critical density $\Gamma_c = (1-b)/(\beta \xi)^2$ and we posed $b = \beta \bar{J}$. While fitting, an extra, province-dependent, parameter referred to as $a$, has to be introduced in order to account for the fact that, due to the scaling between $Y$ and $m$, the former is in principle not bounded. The best-fit coefficients are collected in Fig.~\ref{fig:isto}.
Notably, we checked that these results are in full consistency with the analogous parameters that one would obtain when fitting with the more explicit square root function (\ref{eq:expand}), at least as far as  small values of $\Gamma$ are considered.
In particular, we notice that $\log(a)$ is roughly uniformly distributed along the range $(12, 19)$, suggesting that the extent of exports varies over several orders of magnitude, according to the province considered. On the other hand, $\Gamma_{c}$ looks Poissonian-like distributed and is peaked around $0.003$, suggesting that when immigrants are less than $0.3 \%$ of the whole population inside the province, their presence is ineffective as facilitator of trade with their country of origin.

Note that, through the statistical mechanics route of phase transitions, finding the critical mass is quite simple, while via standard approaches accessing this quantity would be much more complex as $\Gamma_{c}$ is a function of several local variables, as coded in Eq.~\ref{eq:critical}.

\subsubsection{Bilateral trades}

\begin{figure}[h!]
 \includegraphics[width=0.45\textwidth]{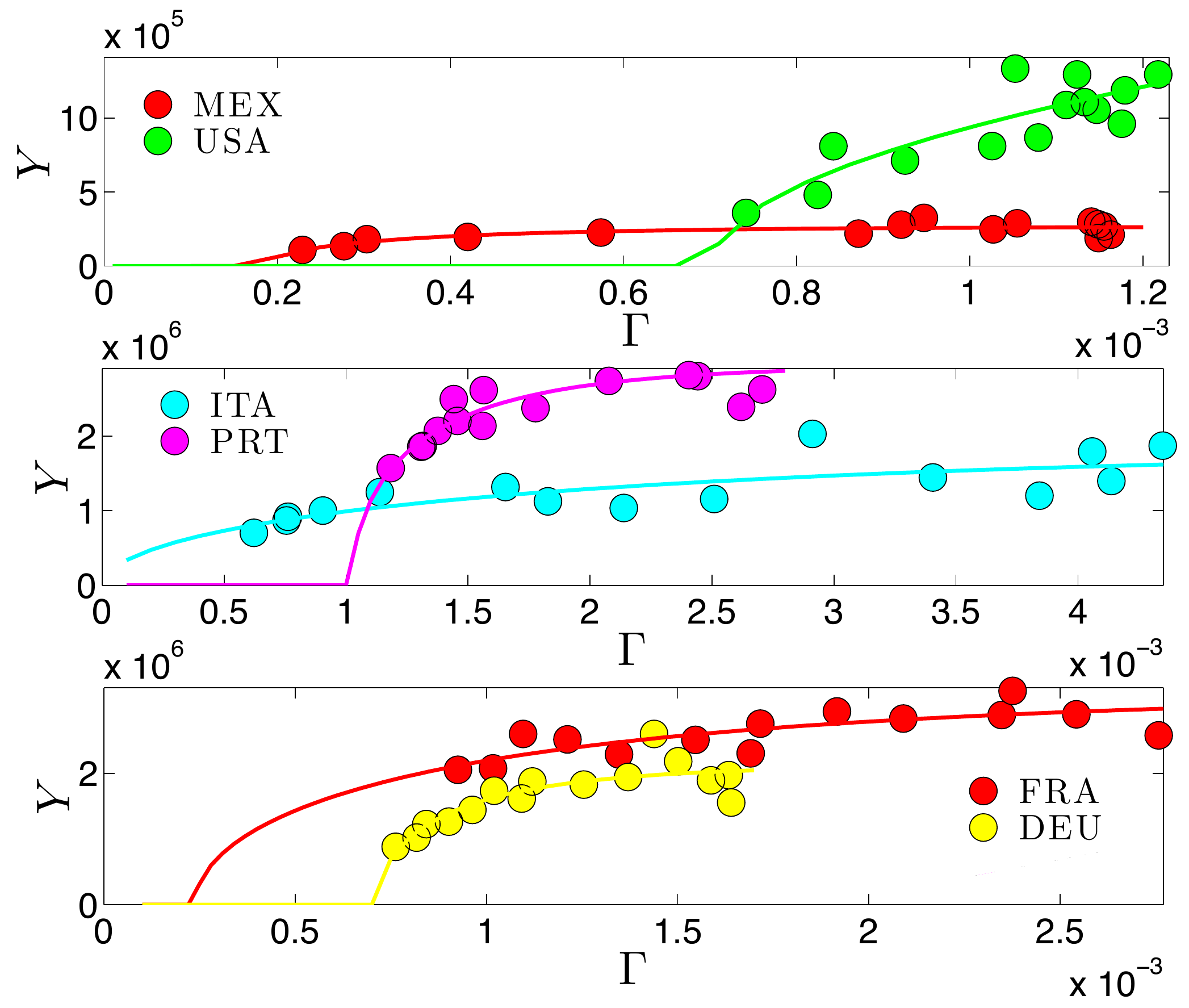}
 \caption{Trades performed by the province of  Madrid  with different foreign countries as a function of the related immigrant density: different countries are depicted in different colours as specified by the legend. Data (bullets) are fitted via Eq.~\ref{eq:tanh2} (solid line). The foreign countries considered are those where $\Gamma$ spans over the largest interval in such a way that fits can be more accurate.}
  \label{fig:Madrid_grandi}
\end{figure}

\begin{figure}[h!]
 \includegraphics[width=0.45\textwidth]{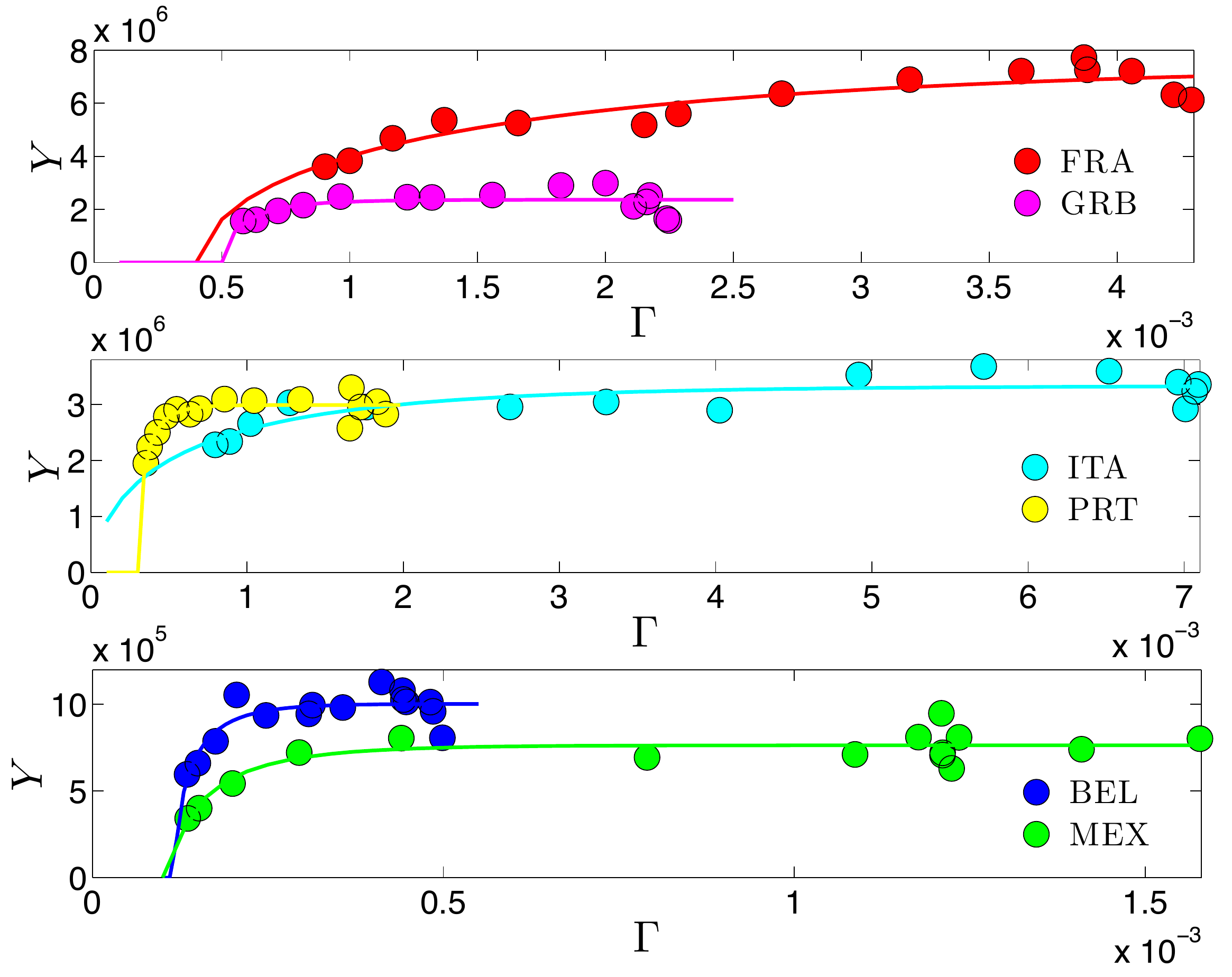}
 \caption{Trades performed by the province of  Barcelona  with different foreign countries as a function of the related immigrant density: different countries are depicted in different colours as specified by the legend. Data (bullets) are fitted via Eq.~\ref{eq:tanh2} (solid line). The foreign countries considered are those where $\Gamma$ spans over the largest interval in such a way that fits can be more accurate.}
  \label{fig:Barca_grandi}
\end{figure}

\begin{figure}[h!]
 \includegraphics[width=0.45\textwidth]{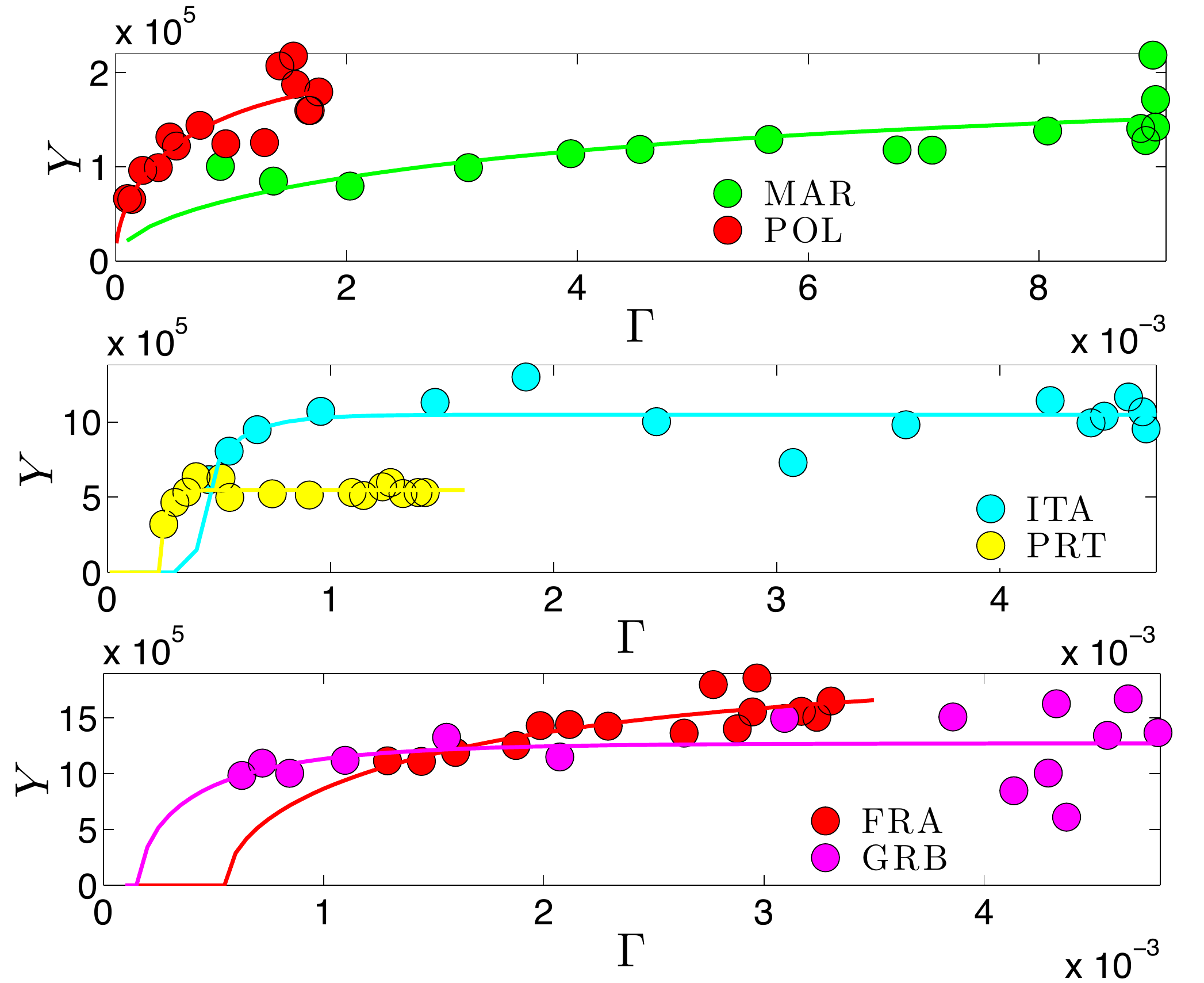}
 \caption{Trades performed by the province of  Valencia  with different foreign countries as a function of the related immigrant density: different countries are depicted in different colours as specified by the legend. Data (bullets) are fitted via Eq.~\ref{eq:tanh2} (solid line). The foreign countries considered are those where $\Gamma$ spans over the largest interval in such a way that fits can be more accurate.}
  \label{fig:Valencia_grandi}
\end{figure}

%It should be stressed that in the results discussed so far nor we  considered the specific country of origin of immigrants neither we analyzed the (related) bilateral trades, that is inspect  trades country-by-country. However, to obtain this finer picture, one should
In order to get a finer picture, and to deepen the possible existence of a country-dependent critical  threshold $\Gamma_c$, we fragment the migrant party into several subsets, each corresponding to a different country of origin and then we analyze the trades $Y_{p,f}$ performed between any province $p$ and any foreign country $f$ as a function of the related fraction of immigrants $\Gamma_{p,f}$. Of course, results are expected to be much more noisy, as we are dealing with considerable smaller datasets and the intrinsic fluctuations are only partially smoothened by the central limit theorem. Nonetheless, it is worth checking whether the previous results are still valid at this less coarse-grained level, and inferring the  country-dependent critical masses. We focused on the three major Spanish cities, namely Madrid (Fig.~\ref{fig:Madrid_grandi}), Barcelona (Fig.~\ref{fig:Barca_grandi}) and Valencia (Fig.~\ref{fig:Valencia_grandi}) and on the foreign countries for which the size of immigrant communities are larger and span along a wide interval in the time window considered, in order to get more accurate and reliable fits.

%As partially expected from the previous argument, when performing such a finer analysis, we find that the theoretical prediction encoded by Eq.~\ref{eq:expand} is no longer fulfilled by small-sized communities (see e.g., the lower panels of Fig.~\ref{fig:MadridCompare} for Madrid and the lower panels of Fig.~\ref{fig:BarcaCompare} for Barcelona), while the largest communities still exhibit a good agreement (see e.g., the upper panel of Fig.~\ref{fig:MadridCompare} for Madrid and the upper panel of Fig.~\ref{fig:BarcaCompare} for Barcelona). We further deepened this in Fig.~\ref{fig:grandi} where exports regarding the twelve countries with largest immigration communities in Barcelona are shown.   We speculate that in the former case the density $\Gamma$ may be smaller than the pertaining critical value and that possible \emph{extensive} thresholds as well as further socio-economical features may also be at work.

\begin{figure}[h!]
 \includegraphics[width=0.45\textwidth]{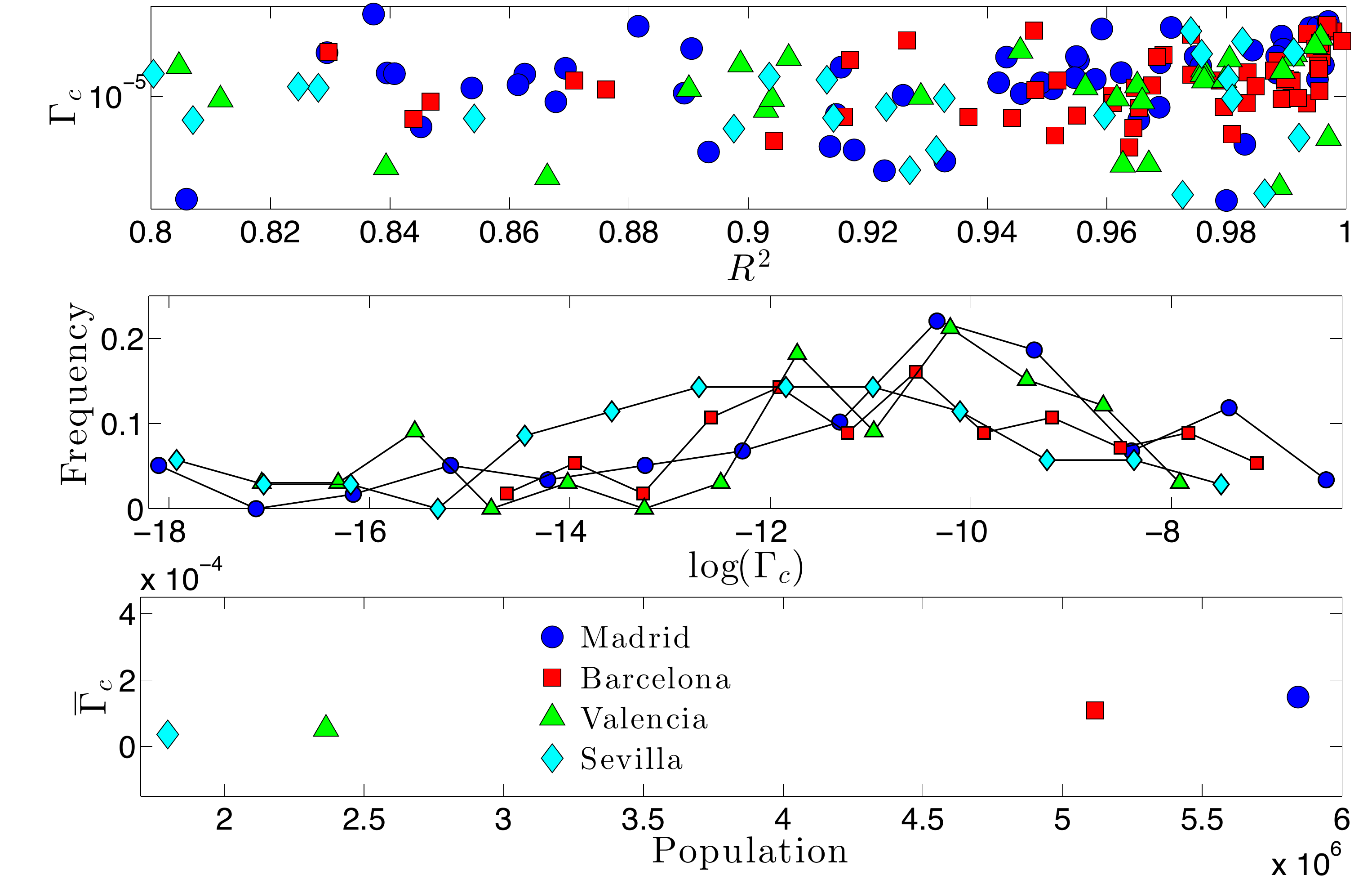}
 \caption{Critical density $\Gamma_c$ estimated by fitting data for trading versus $\Gamma$ according to the theoretical law Eq.~\ref{eq:tanh2}. Analysis are performed for the four largest provinces: Madrid, Barcelona, Valencia, Sevilla depicted in different symbols and colors as explained by the legend. We added Sevilla to confirm with a fourth point the trend depicted by the first three cities. For each available foreign country $f$ we fit the data for trades versus $\Gamma$ according to the tanh law of Eq.~\ref{eq:tanh2} and we derive an estimate for $\Gamma_c$ with the related $R^2$, which are plotted in the topmost panel. The most reliable fits (i.e. $R^2$ close to $1$) suggest that $\Gamma_c$  are scattered around $10^{-5}$, similarly for the four provinces analyzed. Focusing on estimates corresponding to $R^2>0.85$, we build the histogram of $\Gamma_c$ (shown in the middle panel) and calculate the arithmetic average to get  $\overline{\Gamma}_c$, which is plotted in the bottom panel as a function of the population of the related province.
%The error bar denotes the standard deviation stemming from the average operation.
%A close inspection to the cases yielding large $R^2$ and $\Gamma_c \approx 0$ suggests that they typically correspond to North-Eastern Europe countries (i.e., Latvia %for Madrid; Bulgaria, Czech Republic, Romania for Barcelona; Lithuania, Romania, Sweden for Valencia.)
}
  \label{fig:Gamma}
\end{figure}

By fitting data according to Eq.~\ref{eq:tanh2} we derive estimates for $\Gamma_c$ which, in general, depend on both $p$ and $f$, as shown in Fig.~\ref{fig:Gamma}. In particular, $\Gamma_c$ follows a distribution peaked around $\overline{\Gamma}_c \approx 10^{-5}$, that is consistent with the previous value $\sim 3 \cdot 10^{-3}$ as migrants come from $O(10^2)$ different countries.
%This indicates that the existence of a non-null critical value for $\Gamma$ may stem from social (e.g., host and birth country display close cultures, languages, habits  which facilitate integration), economical (e.g., special international agreements) or geographic (e.g., cheap transportation) reasons, just to cite a few.

We finally notice that $\overline{\Gamma}_c$ seems to slightly vary with the size of the hosting population, consistently with expected finite size effects.

\begin{figure}[bt]
\includegraphics[width=0.45\textwidth]{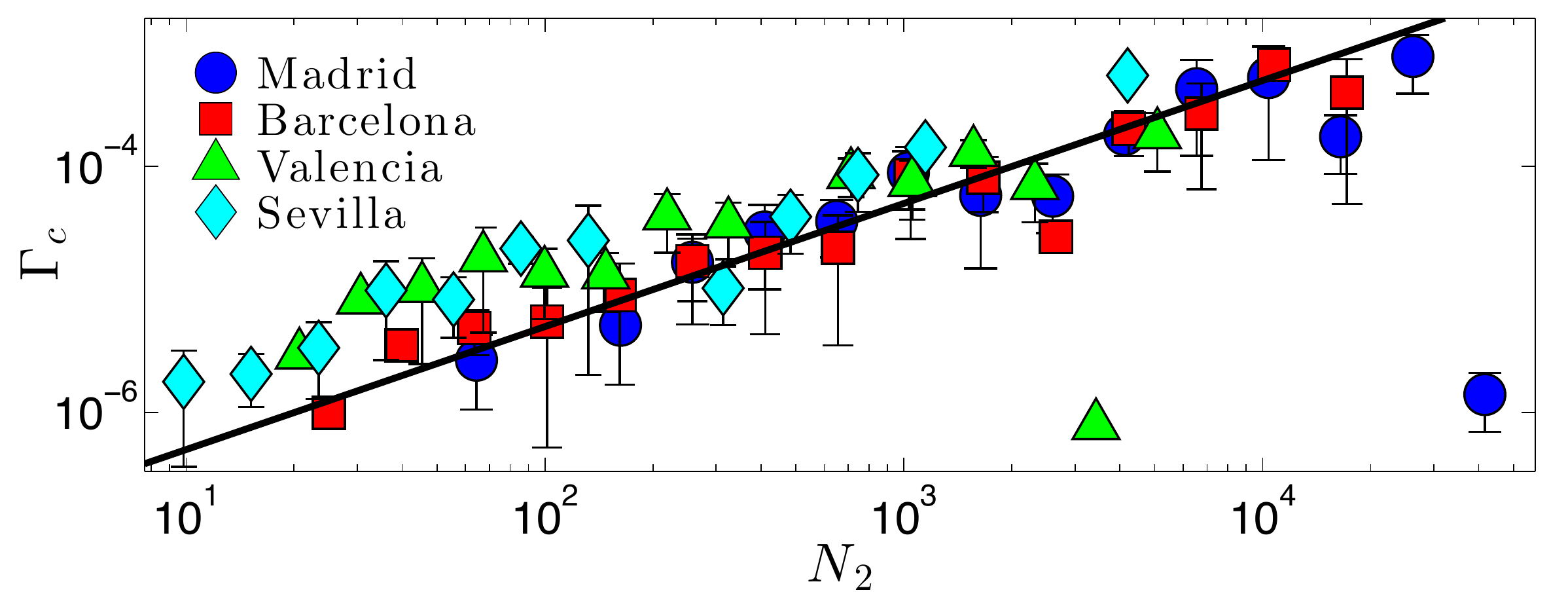}
\caption{The values of $\Gamma_c$, obtained by fitting data for trades between $p$ and $f$ are related to the size $N_2$ of the community of migrants hailing from $f$ and resident in $p$. Here we focused on the four largest provinces and for each we show data stemming from a binning procedure, in such a way that the error bars represents the standard deviation for the data points pertaining to the same bin. }
  \label{fig:lanove}
\end{figure}

Lastly, we checked that there is a clear correlation between the critical value $\Gamma_c$, obtained for trades between $p$ and $f$, and the size $N_2$ of the community of migrants hailing from $f$ and resident in $p$. This linear correlation is confirmed for the four largest cities  we analyzed in detail (i.e. Madrid, Barcelona, Valencia, Sevilla) as shown in Fig.~\ref{fig:lanove}.

%Also, for Barcelona, $\bar{\Gamma}_c \sim 10^{-5}$ and, consistently, according to Fig.~\ref{fig:Barca_grandi} values of $\Gamma \lesssim 10^{-5}$ correspond to too small values of migrant density for the square root law to emerge. The critical value obtained when looking at the overall migrant density, without distinguishing the birth country, was $\sim 10^{-3}$ which seems consistent with the value $\sim 10^{-5}$ obtained through the finer analysis as the number of birth countries considered is $\sim 10^2$.

\subsubsection{The relation between migrants and products diversification}

Having proved that the amount of trades is positively influenced by migration, we still have to check that also the diversification of exports is enhanced, namely, that migration plays a significant role in the modern theory of Economical Complexity.

In order to keep this analysis as simple as possible, we do not deal with recent {\em complexity measures} \cite{pietronero1,pietronero2,pietronero3}, but we follow the simplest possible route (leaving for future works possible improvements).

The export portfolio of a province is composed of products and destinations. That is, a province can export several products to a single destination or export the same product to several destinations. Thus, the basic unit in the export portfolio is a product-destination pair. We define $K$ as the total number of product-destination pairs in the export portfolio of a province.
%$K$ is the simplest quantifier for diversification.
Products are defined using the $HS1996$ product classification \footnote{We excluded "special" product categories ($HS98$ and $HS99$) from COMTRADE database}. Destinations are defined as countries with more than 1 million population in $2010$. There are $4507$ products and $154$ countries, so the total number of product-destination $K$ pairs is $694078$.

To account for the distribution of export sales across product-destination pairs, we use the export share of each product-destination pair in total export value so to capture the relative importance of each pair for exports. The Herfindahl index $N_H$ \cite{index} is a simple calculation of concentration of exports that uses such export shares: the larger the number $N_H$, the more concentrated (less diversified) the export portfolio of the province is. Therefore, if migrants do really contribute to diversification of exports, we should expect a negative correlation between $N_H$ and $\Gamma$. More precisely, the $N_H$ index is calculated as
\be
N_H = \sum_{i=1}^K \left(\frac{x_i}{X} \right)^2,
\ee
where $x$ is the value of export in product-destination $i$ and $X$ is the total value of exports.
One can further normalise $N_H$ to get an index $n_H$ whose values lie between $0$ and $1$.
\begin{figure}[bt]
\includegraphics[width=0.475\textwidth]{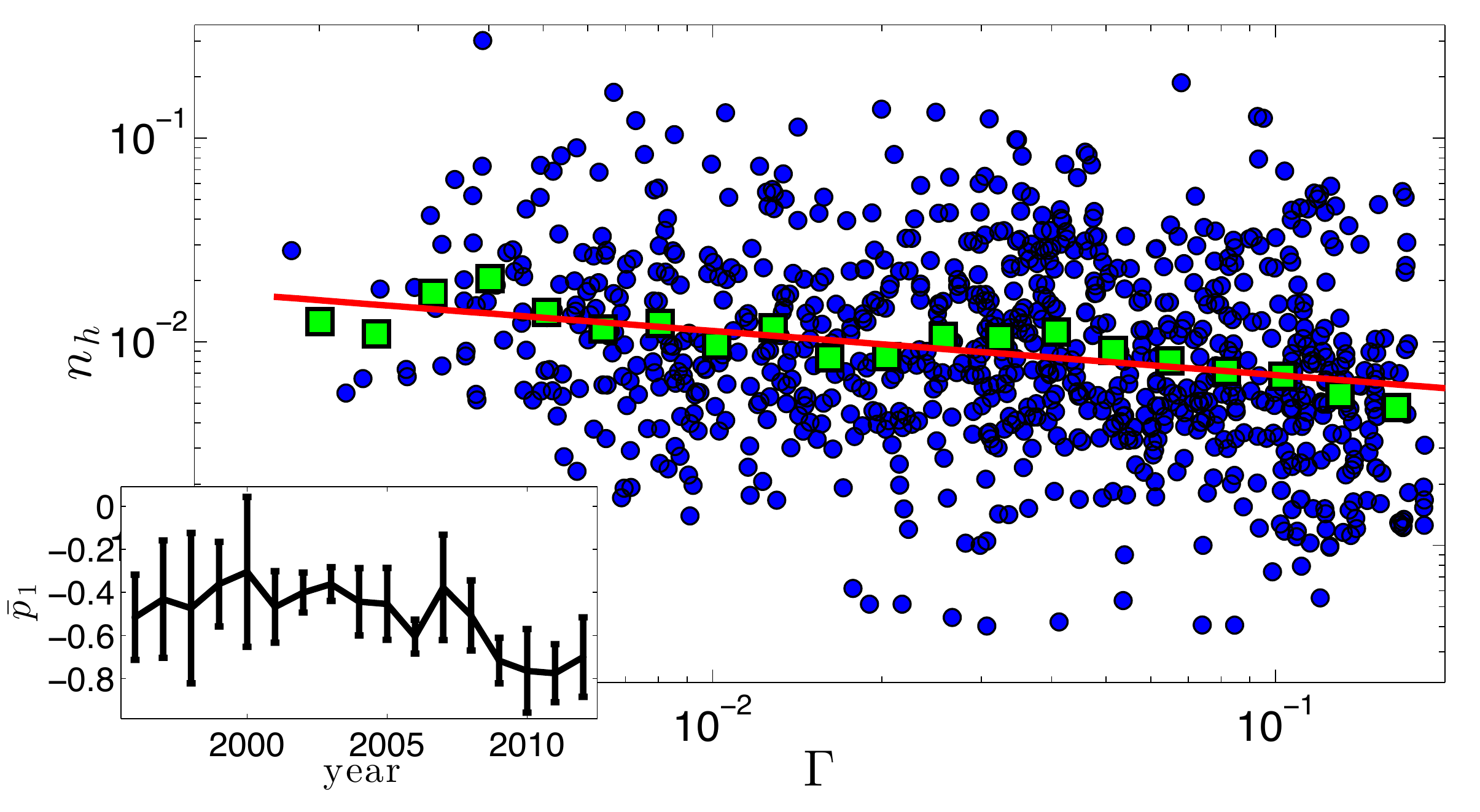}
\caption{ In the main plot bullets represent the value of  the normalized diversification index $n_h$ for different province and different years as a function of $\Gamma$. Green squares represent binned data and the solid red line is the related best fit. This is a linear curve (in log-log scale) $y = p_1 x +p_2$, with $p_1 =  -0.21 \pm 0.01$ and $p_2= -5.47 \pm 0.01$. The fitting has also been performed for data of $n_h$ pertaining to any single province and any single year, hence obtaining $p_1(y,p)$. These values have been averaged over the provinces to get $\overline{p_1}(y)$ which is shown in the inset (the line is a guide for the year). This plot shows that the monotonicity of $n_h$ with respect to $\Gamma$ (i.e. $p_1 <0$) is robust with respect to the year; the same holds even when we average over the year, namely it is robust with respect to the province.
%(Bottom) Positive correlation between the growth in the number of firms in a given province and the relative percentage of migrants; each panel corresponds to a different province, retaining the same legend of Fig.~\ref{fig:esempio}.
}
 \label{fig:main-diversification}
\end{figure}
Results are shown in Fig.~\ref{fig:main-diversification} where the negative correlation between $n_h$ and the percentage of migrants within the province is manifest. Thus, {\em at least for small percentages}, that is $\Gamma \sim \gamma$, there is a positive correlation between export portfolio diversification and the density of migrants in a particular province. We can therefore derive that migrants act as facilitators of trade by reducing international transaction costs.
%\newline
%Note that each export transaction is invoiced by an exporting firm to one foreign firm, hence an increase in the number of export transactions to one country may reflect more exporting firms, thus new trading relations (extensive margin), or higher frequency in trading relations between already existing trading partners (intensive margin): by reducing the cost of making business in the country of origin of migrants, their community living in the host country tacitly increases the number of transactions from the province itself to the country of origin \cite{francisco1,francisco2}.
%Restricting the scenario among the two possibilities, an increase in the intensive margin or an increase in the extensive margin, the latter seems to be the solely in agreement with a self-averaging theory, as the one here developed: however, it should be noticed that our approach -at the refined level of bilateral trades- reproduces with remarkably accuracy only  a part of the data,  and not the totality. It could still be that the remaining part not best fitting within our theory is driven by mechanisms involving the intensive margin and, thus, by assumption, escapes from the scenario exploited in the paper.

\subsection{PART TWO: Insights from Graph Theory}

The interaction between natives and immigrants was described in terms of a bipartite graph (see Fig.~$3$, left panel). The statistical mechanics analysis shows that if the local agents $i$ and $j$ both interact with some foreign-born individual $\mu$, i.e. $\xi_i^{\mu},\xi_j^{\mu} \neq 0$, then the agents $i$ and $j$ can be thought of as directly interacting via an effective coupling $\tilde{J}_{ij} \sim \sum_{\mu} \xi_i^{\mu} \xi_j^{\mu}$ (see Fig.~$3$, right panel and Eq.~\ref{eq:hebb}).
We now focus on such emergent network, referred to as $\mathcal{G}$ and, through calibration with available data, we try to infer information for the test case of Spain.

% as reported in details in Appendix B.
%Here we describe its main properties.

\subsubsection{A glance at the theory}
The topological properties of $\mathcal{G}$ have been formerly mathematically investigated in \cite{Agliari-EPL2011,Barra-JStat2011,Barra-PhysA2012} and here we review the main points.

A global characterization of the graph $\mathcal{G}$ can be attained in terms of the average link probability $p$: considering a generic couple of nodes, say $i$ and $j$, keeping a mean-field perspective, we can write
\ba
%p &=&  1 - \prod_{\mu=1}^{N_2} \left( 1 - \frac{\xi_i^{\mu}}{N^{\theta}} \frac{\xi_j^{\mu}}{N^{\theta}} \right)  \\
%&=& 1 - \left( 1 - \frac{\xi_i^{\mu}}{N^{\theta}} \frac{\xi_j^{\mu}}{N^{\theta}} \right)^{N_2} = 1 - \left( 1 - \frac{\xi^2}{N^{2\theta}}  \right) %^{\gamma N},
\label{eq:p1}
p &=&  1 - \prod_{\mu=1}^{N_2} \left[ 1 - \mathbb{P}(\xi_i^{\mu}=1)\mathbb{P}(\xi_j^{\mu}=1) \right]   \\
\label{eq:p2}
&=&  1 - \left( 1 - \frac{\xi^2}{N^{2\theta}}  \right) ^{\gamma N},
\ea
where in Eq.~(\ref{eq:p1}) the term in the square brackets represents the probability that the contribution $\xi_i^{\mu}\xi_j^{\mu}$ in the sum (\ref{eq:hebb}) is equal to zero and the product over $\mu$ returns the probability that all entries $\mu =1 ,..., N_2$ are null such that, finally, the complementary of this quantity provides the probability that at least one entry is non-null, that is, that $\tilde{J}_{ij}>0$; in Eq.~(\ref{eq:p2}) we used the homogeneity of pattern entries (\ref{eq:pattern}) and the definition of $\gamma$ (\ref{eq:gamma}).
\newline
The average degree of $\mathcal{G}$ therefore reads as $\bar{d} = p N_1$.

Now, as $\theta$ and $\xi$ are tuned, the emerging graph can range from fully-connected to completely disconnected \cite{Agliari-EPL2011,Barra-JStat2011}. From a mean-field perspective, we can distinguish the following topological regimes:
\begin{itemize}
\item  $\theta <1/2$, \; $p \rightarrow 1$, \; $\bar{d} \rightarrow N$  \; $\Rightarrow$  Fully connected (weighted) graph.\\
\item  $\theta =1/2$, \; $p \sim 1- e^{-\xi^2 \gamma}$, \;  $\bar{d} = \mathcal{O}(N)$ \; $\Rightarrow$ Linearly extensive degree.\\
\item  $1/2 < \theta <1$, $p \sim \xi^2 \gamma  N^{1-2 \theta}$,  $\bar{d} = \mathcal{O}(N^{2(1 -  \theta)})$ $\Rightarrow$  Extreme dilution regime: $\lim_{N \rightarrow \infty}  \bar{d}^{-1} = \lim_{N \rightarrow \infty} \bar{d}/N = 0$.\\
\item  $ \theta = 1$, \; $p \sim \xi^2 \gamma/N$, \; $\bar{d} = \mathcal{O}(N^{0})$ \; $\Rightarrow$  Sparse (weighted) graph;  $\xi^2 \gamma = 1$ corresponds to the percolation threshold.
\end{itemize}
Summarizing, large values of $\theta$ determine a disconnected graph with vanishing average degree.
Therefore, $\theta$ coarsely controls the connectivity regime of the network, while $\xi$ and $\gamma$ allow a finer tuning.

As the graph $\mathcal{G}$ is meant to describe the mutual interactions among the decision makers inside a society, it is worth investigating whether it also exhibits any of the small-world hallmarks.
Indeed, as shown in \cite{Agliari-EPL2011,Barra-JStat2011,Barra-PhysA2012}, this is the case: for instance, in the proper parameter range, $\mathcal{G}$ is shown to display a small diameter and a high clustering coefficient.
%
%We find that $\mathcal{G}$ displays a high frequency of triangles and the mean-shortest
%path length among two nodes grows logarithmical with $N_1$. While these statements are extensively proven in  \cite{Agliari-EPL2011,Barra-JStat2011}, here we provide simple, intuitive arguments to see this.
%As for the diameter, a logarithmic growth is known to be a common property of random graphs (provided that they do not display any pathological inhomogeneity) \cite{Albert-RevModPhys2002,Newman-SIAM2003}  [20,21], and is therefore implicitly satisfied also by the graph under study. As for the clustering coefficient, we notice that
%
In fact, the definition in Eq.~\ref{eq:hebb} (i.e. the Hebbian kernel) implicitly endows couplings with ``transitivity'': if $i$ and $j$ are connected as they share acquaintances among immigrants, and the same holds for $i$ and $z$, then $j$ and $z$ are also likely to share any acquaintance. Otherwise stated, interactions based on \emph{sharing} (i.e., matching non-null entries) intrinsically generate a clustered society.

%\subsubsection{The role of weak ties}

Up to now we just focused on the bare topology, yet the graph $\mathcal{G}$ is weighted and we can wonder whether, even from this perspective, the graph exhibits typical features of social networks.

%As for couplings, we can still detect ``modes'', each characterized by $J_{\rho}$ representing the average strength for links stemming from a node  associated to a string with $\rho$ non-null entries. While $J_{\rho}$ provides a measure of the local ``field'' seen by a single node, a global description can be attained by the overall average coupling $\bar{J} \equiv \sum_{\rho} J_{\rho} P_1(\rho; a,L)$, taken over the whole graph; the two quantities are related via $J_{\rho} = \sqrt{\bar{J}}\rho/L$ \cite{nostro}. As we will see in the next section, despite the self-consistence relation (more sensible to local
%conditions) is influenced by $\sqrt{\bar{J}}$, the critical
%behavior occurs at $\beta_c = \bar{J}^{-1}$, consistently with
%a manifestation of a collective, global effect.
%
%By looking in more detail at the coupling distribution holding in the thermodynamic limit and for values of $a$ determined by Eq.~\ref{eq:scaling}, we find that, for $1/2 < \theta \leq 1$, nodes are pairwise either non-connected or connected due to one single matching among the relevant strings.
%For $\theta =1/2$ this still holds when $\alpha \gamma^2 /4 \ll 1$, which corresponds to a relatively high dilution regime, otherwise some degree of disorder is maintained.
%On the other hand, for $\theta < 1/2$, while topological disorder is lost, disorder on couplings is still present. However, for $\theta = 0$ and $\gamma = 2$, the coupling distribution gets peaked at $J=L$ and, again, disorder on couplings is lost \cite{nostro}.
%

In particular, according to the {\em strength of weak ties} theory by Granovetter \cite{Granovetter-1973, Granovetter-1983}, the degree of overlap of two individuals' neighbourhood varies directly with the strength of their
tie to one another. If the two individuals are acquaintances (rather than close friends),
there is little overlap.
%Such weak ties turn out to be of crucial importance as they are more likely to convey non-redundant information.
Consistently, in the graph $\mathcal{G}$ weak ties connect individuals sharing a small number (possibly only one) of connections in the immigrant community.

%Under assumption (\ref{eq:pattern}),Agliari-PRE2011

Finally, as shown in \cite{Barra-PhysA2012,Agliari-PRE2011,callaway,barabasi,marti,vespignani}, weak ties also turn out to be crucial in order to maintain the network connected: by cutting (a relatively small number of) weak link the network gets fragmented into several components.

%In order to quantitatively investigate this feature, we diluted the graph $\mathcal{G}$ following threes different procedures: $i.$ by cutting links associated to a weight progressively smaller, $ii.$ by cutting links associated to a weight progressively larger, $iii.$ by cutting links extracted randomly. We then compared the size of the largest component $S$ and its behaviour with respect to the fraction of links deleted finding that when weak ties are removed first the giant component displays the (early) fastest decrease. On the other hand, when strong ties are removed first the giant component displays the (early) slower decrease. This suggests that weak ties are crucial for the network to remain connected \cite{Barra-PhysA2012,Agliari-PRE2011} thus allowing better information spreading \cite{vespignani}.

\subsubsection{Inferring the topological properties}

The parameters into play are $\gamma, \theta$ and $\xi$; their values determine the topology of the emergent network and are also expected to affect the growth of trades (see e.g. Eq.~\ref{eq:expand}).
Let us now try to estimate them starting from empirical data.

\begin{figure}[tb]
 \includegraphics[width=0.45\textwidth]{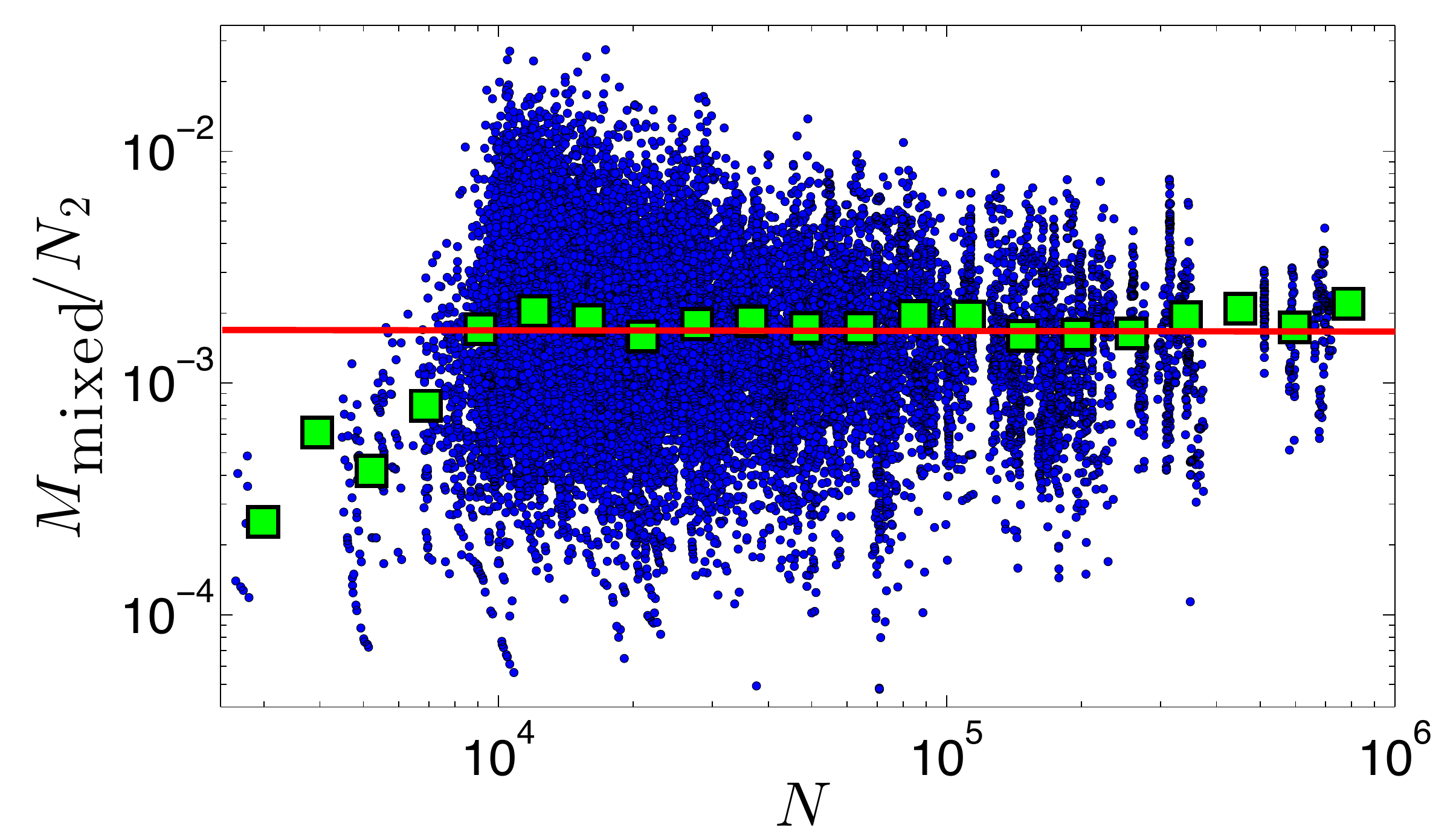}
 \caption{Data on mixed marriages for each province along the years 1998-2012 are drawn from the local offices of Vital Records and Statistics (Registro Civil) and divided by the related size $N_2$ of the immigrant community. Raw data (blue bullets) are properly binned (green squares) to highlight the effective behaviour with respect to the overall size $N$ of the related province. The red line shows the lack of dependence on $N$, for $M_{\textrm{mixed}}/N_2$, in the large $N$ limit.}
 \label{fig:ansatz}
\end{figure}

As for $\theta$, we can derive it through an indirect measure:  we expect that the number of links between locals and immigrants is lower bounded by the number of mixed marriages $M_{\textrm{mixed}}$. In fact, a mixed marriage yields, in general, several ``mixed acquaintances'' between the family members and the friends of the two parties.
In complete generality, the probability of mixed marriage $p_{\textrm{mixed}}$ also scales with $N$, that is  $p_{\textrm{mixed}} \sim N^{- \tilde{\theta}}$, with $\tilde{\theta} \geq \theta$, therefore, we can write
\begin{equation}
M_{\textrm{mixed}} \sim N_1 \times N_2 \times p_{\textrm{mixed}}  \sim N_1 \times N_2 \times \frac{\xi}{N^{\tilde{\theta}}}
\end{equation}
from which
\begin{equation}
\frac{M_{\textrm{mixed}} }{ N_2} \sim  N^{1 - \tilde{\theta}}.
\end{equation}
Mixed marriages in Spain have been thoroughly investigated in \cite{Barra-ScRep, Agliari-NJP} and from those data we can fit the ratio $M_{\textrm{mixed}}/N_2 \sim N^{1- \tilde{\theta}}$ inferring an estimate for $\tilde{\theta}$.  As shown in Fig.~\ref{fig:ansatz}, the number of normalized mixed marriages is roughly constant with respect to $N$, that is $\tilde{\theta} \approx 1$.
As a consequence, $\theta \leq 1$.

Now, a value of $\theta$ strictly smaller that $1$ would imply that the number of connections between the two parties grows indefinitely with $N$ (or, analogously, with $N_1$ or $N_2$), and this is certainly not realistic (it would imply infinite energy in order to sustain such a network and the linear extensivity of its related thermodynamics would breaks down). Thus, the experimental argument for the lower bound coupled with the theoretical argument for the upper bound implies $\theta =1$ (as intuitive).

\begin{figure}[h!]
 \includegraphics[width=0.45\textwidth]{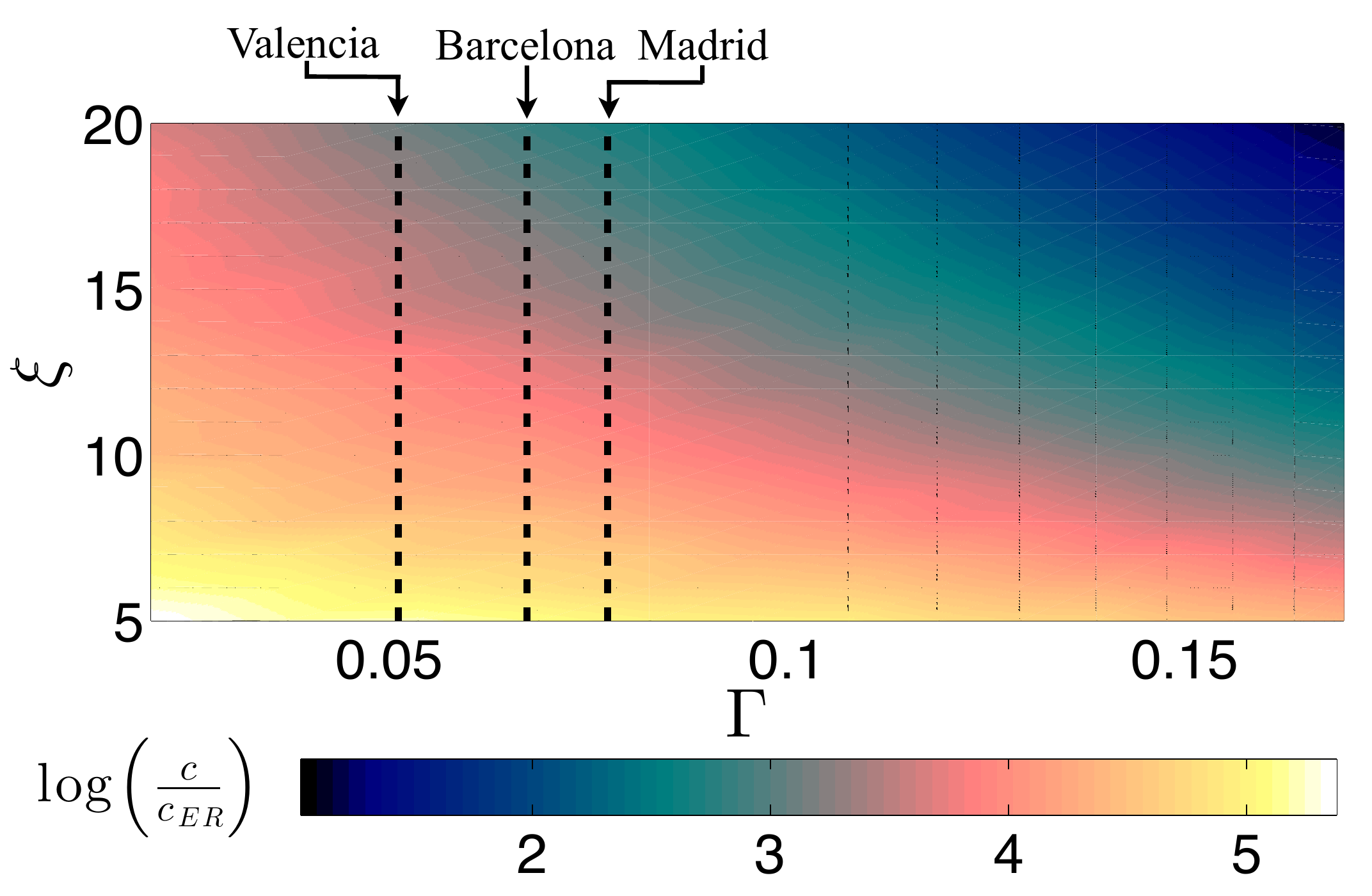}
 \includegraphics[width=0.45\textwidth]{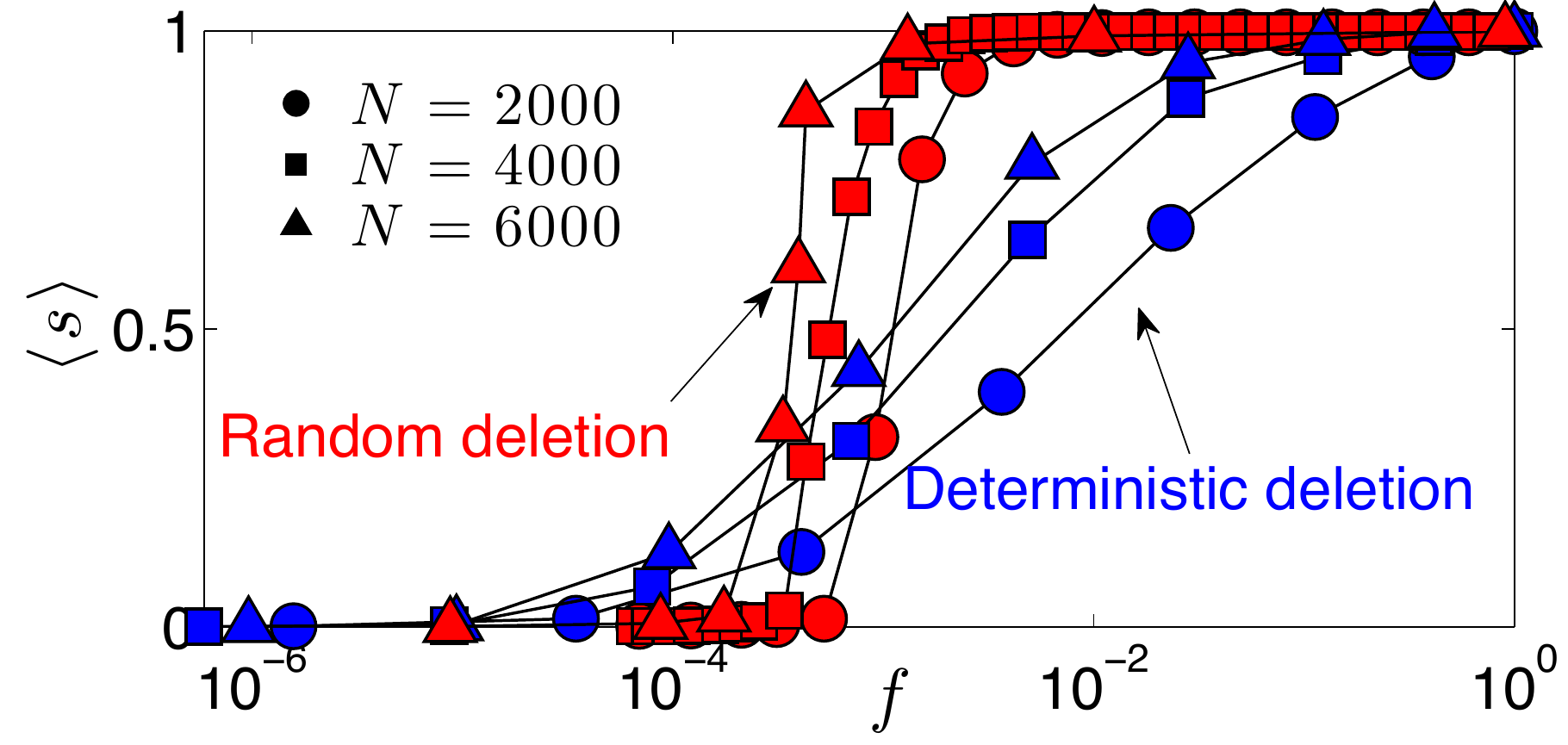}
 \caption{Upper panel: Comparison between $c_{\textrm{ER}}$ and $c$, as a function of $\Gamma$ and $\xi$. The empirical values of $\Gamma$ for the test-case provinces are also shown.
 Lower panel: Average size $\langle s \rangle$ of the giant component
obtained by bond-percolating $\mathcal{G}$, being $1-f$ the fraction of
links deleted. Two processes are compared: random dilution
(links to be deleted are extracted randomly) and deterministic
dilution (links to be deleted are chosen starting from
those with lower weight). Remarkably, in the latter case, by
deleting the weakest links corresponding to a small fraction
$1-f$ of the overall links, the graph already gets fragmented in
several components. See \cite{Barra-PhysA2012} for more details. }\label{fig:clustering}
\end{figure}

Finally, we need to estimate $\xi$. According to Eq.~\ref{eq:pattern}, and having fixed $\theta=1$, $\xi$ represents the average number of local acquaintances displayed by an immigrant. In our analysis we bound $\xi$ in between $1$ (we expect that any immigrant has at least one link with the local community) and $20$: there are several sociological studies trying to estimate the average number of acquaintances (familiars and/or friends) of a member of societies. In particular, in \cite{Estudio,HumanOrganization} this analysis is performed in Spain finding that this number is $\xi = \mathcal{O}(10)$, similarly to other European countries.

According to these estimates for $\theta, \xi$, and $\gamma$ we expect a sparse graph and we can check whether the emergent graph is indeed clustered.
In Fig.~\ref{fig:clustering} (upper panel) we show the ratio between the average clustering coefficient $c(\gamma,\xi)$ measured in a numerical realization of $\mathcal{G}$ and the clustering coefficient $c_{\textrm{ER}}$ of an analogous Erd\"{o}s-R\'enyi graph.
More precisely, $c(\gamma,\xi)$ is measured as a function of $\xi$ and $\gamma$, varied within the ranges empirically detected as described above; for each choice of parameters we can derive an average degree $\bar{d}$ which is used to estimate $c_{\textrm{ER}}$, namely $c_{\textrm{ER}}= \bar{d}/N_1$. As long as $c / c^{\mathrm{ER}} >1$, $\mathcal{G}$ is highly clustered and this occurs in a wide region of the plane $(\Gamma,\xi)$, especially in the region of high dilution: in the parameter range considered the graph $\mathcal{G}$ turns out to be small world.

Lastly, we address the problem of the existence of {\em weak ties} within the network generated by the Hebbian rule (eq. $5$). To this task, we have numerically built over-percolated networks at various sizes and then, for each sample, we performed two types of dilution: the former is purely random, namely we delete a fraction of links extracted according to a uniform distribution, the latter is deterministic, namely we delete links selecting those corresponding to the weakest coupling.  We can then compare the results (shown in Fig.~\ref{fig:clustering} , lower panel). If weak ties effectively play a crucial role in keeping different communities connected together, then the deterministic percolation should break the giant component first (i.e. at higher values of network's connectivity) as this is the case, hence, at least numerically, we definitely confirm that these Hebbian networks are small worlds.

\section{Conclusions and outlooks}

The recent results of Hidalgo, Klinger, Barabasi and Hausmann \cite{barabasi2,barabasi3}, as well as those by Pietronero, Caldarelli, Gabrielli and their coworkers \cite{pietronero1,pietronero2} play as breakthrough in the modern theory of Economical Complexity: while classical economic theories prescribed specialization in the industrial production of most developed countries, their investigations clearly show that nowadays the production of such countries are actually extremely diversified. However, in these papers, how diversification affects international trades is not deepened and this is the goal of the present work.

Our approach is framed within the scaffold of Statistical Mechanics, a well consolidated stochastic tool in Theoretical Physics that aims to detect emergent and collective behaviors sharing attributes over the details, and it is supported by extensive data analysis for the test-case of Spain.
The resulting theory plays as a new dowel in this modern mosaic of Economical Complexity, shedding lights on the way diversification  of exports is achieved due to a continuous swarming of natives and migrants in interaction. These exchanges of information are fundamental to allow firm's holders to leverage transactional costs thus tacitely allowing a larger basin of firms to appear on the international market.

From a practical economical perspective, our results suggest the existence of a (eventually very small) critical threshold $\Gamma_c$ in the percentage of migrants present in the host community before a boost in international trading is achieved, as well as a saturation effect,
%coded in the narrowness  of the hyperbolic tangent, see eq. \ref{main-part1}
in agreement both with the Chaney distorted gravity scheme as well as with recent non linear models by Egger et al. \cite{egger} and the (related) pioneering suggestions of Gould \cite{gould}.

It is worth highlighting that, through an analogy with phase transitions, we can quantitatively find the probability distribution of the critical threshold, that, when considering migrant's from all over the world as a whole,  is Poissonian-like distributed with peak at $0.3\%$ of the whole population,
%(see Fig.~\ref{fig:isto}, right panel),
while, when considering migrants from a specific country, decreases to values $\Gamma_c \sim 10^{-5}$, whose scaling is in agreement with the observation that trading nations are $\sim 10^2$:

Summarizing, under the assumption of not so large migrant's densities, the effect of immigrant's networking on exports is always significant, robust and stable across goods. Indeed, we can safety state that migrants play a significant role in the modern theory of Economical Complexity.

Further outlooks may cover more complex measures of product's complexity in order to better tackle the outlined (indirect) influence of migrants to the global market and other nations should be considered beyond Spain to give more ground to the theory as a whole.

\section*{Acknowledgments}
The authors are indebted with Luciano Pietronero for very interesting discussions on Economical Complexity and with Francesco Guerra for enlightening discussions on Statistical Mechanics.
\newline
This work is supported by Gruppo Nazionale per la Fisica Matematica (GNFM-INdAM) via the grant Progetto Giovani Agliari2014.
\newline
AB is grateful to LIFE group (Laboratories for Information, Food and Energy) for partial financial support through  {\em programma INNOVA, progetto MATCH}.

\appendix

\section{Statistical- mechanics analysis}
Let us introduce the set of order parameters $\{m_p^{\mu}(\sigma_p)\}_{\mu=1}^{N_2}$ (where the subscript $p$ refers to a given province $p$) as
\be
m_p^{\mu}(\sigma_p)=\frac{1}{C}\sum_{i=1}^{N_1}\xi^{\mu}_i\sigma_{(i_p)},
\ee
where $C$ normalizes with respect to the expected number of non null entries, namely
\be
C=N_1\mathbb{P}(\xi^{\mu}_i=1)=N_1\xi N^{-\theta}=\xi(1-\gamma)N^{1-\theta}.
\ee

Therefore, $m_p^{\mu}(\sigma)$ is the average will in international trading for Spanish people living in the province $p$ that share the knowledge of the migrant $\mu$.
\newline
For the gauge-like symmetry of the model, clearly $\left\langle m^{\mu}(\sigma)\right\rangle=m$ for each $\mu=1,\cdots, N_2$.  Further, as in the dilution regime of empirical interest each decision maker $\sigma_i$  is linked with (at least) one stranger $\mu$, we have that $\left\langle m(\sigma)\right\rangle=\left\langle m^{\mu}(\sigma)\right\rangle=m$, i.e. $m$ is the averaged predisposition of the whole host community in international trading.
This is because
\begin{eqnarray}\small
\left\langle m(\sigma)\right\rangle&=&\left\langle \frac{1}{N_1}\sum_{i=1}^{N_1}\sigma_i\right\rangle=\left\langle \sigma_i\right\rangle\nonumber \\ \small
\left\langle m^{\mu}(\sigma)\right\rangle&=&\left\langle \frac{1}{C(N)}\sum_{i=1}^{N_1}\xi^{\mu}_i\sigma_i\right\rangle=\frac{N_1}{C(N)}\left\langle \xi^{\mu}_i\sigma_i\right\rangle\nonumber \\ \small
&=&\frac{N_1}{C(N)}\mathbb{P}(\xi^{\mu_i}=1)\left\langle \sigma_i\right\rangle=\left\langle \sigma_i\right\rangle.
\end{eqnarray}
In terms of these new order parameters we can write
\be
Z=\sum_{\sigma}e^{\frac{\beta^2\xi^2(1-\gamma)^2}{2}\sum_{\mu=1}^{N_2}m_{\mu}^2(\sigma)}
\ee
that can be evaluated straightforwardly with a standard saddle point argument:
\begin{eqnarray}\small
Z&=&\int\prod_{\mu}dm_{\mu}\int\prod_{\mu}d\hat{m}_{\mu}e^{\frac{\beta^2\xi^2(1-\gamma)^2}{2} }\\  &\cdot& e^{\sum_{\mu=1}^{N_2}m_{\mu}^2}e^{i\sum_{\mu=1}^{N_2}\hat{m}_{\mu}m_{\mu}}\sum_{\sigma}e^{-i\sum_{\mu=1}^{N_2}\hat{m}_{\mu}m_{\mu}(\sigma)}\nonumber \\ \nonumber \small
&=&\int\prod_{\mu}dm_{\mu}\int\prod_{\mu}d\hat{m}_{\mu}e^{\frac{\beta^2\xi^2(1-\gamma)^2}{2}\sum_{\mu=1}^{N_2}m_{\mu}^2}\\ \small \nonumber &\cdot& e^{i\sum_{\mu=1}^{N_2}\hat{m}_{\mu}m_{\mu}}  e^{N_1\log 2+ N_1\mathbb{E}\log\cosh(\frac{-i\sum_{\mu=1}^{N_2}\hat{m}_{\mu}\xi^{\mu}}{\xi(1-\gamma)N^{1-\theta}})}.
\end{eqnarray}
As well known, only the dominant term contributes to the free energy in the large $N$ limit, hence
\begin{eqnarray}
Z&=&\int\prod_{\mu}dm_{\mu}\int\prod_{\mu}d\hat{m}_{\mu}e^{N_1\mathcal{A}(\{m_{\mu}\},\{\hat{m}_{\mu}\})} \\ \nonumber
&\sim& e^{N_1 \sup_{\{m_{\mu}\},\{\hat{m}_{\mu}\}}\mathcal{A}(\{m_{\mu}\},\{\hat{m}_{\mu}\})},
\end{eqnarray}
where
\begin{eqnarray}\small
\mathcal{A}(\{m_{\mu}\},\{\hat{m}_{\mu}\})&=&\log 2 + \mathbb{E}\log\cosh(\frac{-i\sum_{\mu=1}^{N_2}\hat{m}_{\mu}\xi^{\mu}}{\xi(1-\gamma)N^{1-\theta}}) \nonumber \\  &+& \frac{\beta^2\xi^2(1-\gamma)^2}{2N_1}\sum_{\mu=1}^{N_2}m_{\mu}^2+\frac{i}{N_1}\sum_{\mu=1}^{N_2}\hat{m}_{\mu}m_{\mu}. \nonumber
\end{eqnarray}
Taking the $\sup$ of $\mathcal{A}$ we get the self-consistent relations of the system
\begin{eqnarray}\small \nonumber
&& \partial_{m_{\mu}}  \mathcal{A}(\{m_{\mu}\},\{\hat{m}_{\mu}\}) = 0  \to  \\ \nonumber
&& \frac{i}{N_1}\hat{m}_{\mu}=-\frac{\beta^2\xi^2(1-\gamma)^2}{N_1} m_{\mu} \to \ -i\hat{m}_{\mu}=\beta^2\xi^2(1-\gamma)^2 m_{\mu}
\end{eqnarray}
\begin{eqnarray}\small \nonumber
&&\partial_{\hat{m}_{\mu}} \mathcal{A}(\{m_{\mu}\},\{\hat{m}_{\mu}\}) = 0 \to \\ \nonumber \small  && \frac{i}{N_1}m_{\mu}=\frac{i}{\xi(1-\gamma)N^{1-\theta}}\mathbb{E}\left[\xi^{\mu}\tanh(\frac{-i\sum_{\nu=1}^{N_2}\hat{m}_{\nu}\xi^{\nu}}{\xi(1-\gamma)N^{1-\theta}})\right],
\end{eqnarray}
that, once solved together, returns the value of the order parameter as a solution of
\be\small
m_{\mu}=\left(\frac{\xi}{N^{\theta}}\right)^{-1}\left\langle \xi^{\mu}\tanh(\beta^2(1-\gamma)\xi N^{\theta-1}\sum_{\nu=1}^{N_2}m_{\nu}\xi^{\nu})\right\rangle_{\xi}.
\ee
Evaluating explicitly the average over $\xi^{\mu}$, we get
\be\small
m_{\mu}=\left\langle \xi^{\mu}\tanh(\beta^2(1-\gamma)\xi N^{\theta-1}(m_{\mu}+\sum_{\nu\neq\mu}^{N_2}m_{\nu}\xi^{\nu}))\right\rangle_{\xi}.
\ee
Now we can look for the solution  $m_{\mu}=m$ for each $\mu$: killing the vanishing term $N^{\theta-1}m$ (that goes to  zero in the thermodynamic limit) we find that $m$ obeys
\be
m=\left\langle \tanh(\beta^2\xi(1-\gamma)\eta m)\right\rangle_{\eta},
\ee
where we defined the random variable $\eta=N^{\theta -1}\sum_{\nu\neq\mu}^{N_2}\xi^{\nu}\sim N^{\theta -1}\sum_{\nu=1}^{N_2}\xi^{\nu}$.

Evaluating the momenta of $\eta$ is straightforward as
\begin{eqnarray}\small
\left\langle \eta\right\rangle &=& N^{\theta-1}N_2\mathbb{E}[\xi^{\nu}]=\gamma \xi ;\nonumber \\ \nonumber \small \small
\operatorname{Var}[\eta]&=&N^{2(\theta-1)}N_2\operatorname{Var}[\xi^{\nu}]=\mathcal{O}(N^{2(\theta-1)}NN^{-\theta})\stackrel{N\to\infty}{\rightarrow} 0.
\end{eqnarray}
This means that in the limit of infinite size, the signal $\eta$ is deterministic and thus we have
\be\nonumber
m=\tanh(\beta^2\xi^2\gamma(1-\gamma)m).
\ee
This formula relates the expected amount of trading firms (and, similarly, the expected volume of international trades) with the fraction $\gamma$ of foreign-born people in the province considered.

The full Hamiltonian (\ref{eq:Hami}) also contains an intra-party interaction term encoded by $\mathbf{J}$, which was not considered in this treatment.
Accounting also for this term would simply imply an additional term $\beta \bar{J} m$ in the argument of the hyperbolic tangent, namely
\be\label{main-part1A}
m=\tanh(\beta \bar{J}m +\beta^2\xi^2\Gamma m),
\ee
where we wrote $\Gamma=\gamma(1-\gamma)$ for simplicity.


\begin{thebibliography}{}

\bibitem{Barra-ScRep} A. Barra,  et al., {\em An analysis of a large dataset on immigration integration in Spain: The statistical mechanics perspective of Social Action}, Sc. Rep. \textbf{4}, 4174, (2014).

\bibitem{Agliari-NJP}  E. Agliari, et al.,  {\em A stochastic approach for quantifying immigrant's interactions},  New J. Phys. \textbf{16}, 103034, (2014).


\bibitem{barabasi2} C.A. Hidalgo, B. Klinger, A.L. Barabasi, R. Hausmann, {\em The Product Space Conditions the Development of Nations}, Science \textbf{317}, 482--487, (2007).

\bibitem{barabasi3} C.A. Hidalgo, R. Hausmann, {\em The building blocks of Economical Complexity}, Proc. Natl. Acad. Sc. USA \textbf{106}, 10570-10575, (2009).


\bibitem{old1} D. Ricardo,  {\em On the Principles of Political Economy and Taxation}, John Murray (1817).

\bibitem{old2} I.L. Flam, M. Flanders, {\em Trade Theory},  MIT Press, Cambridge (1991).

\bibitem{pietronero1} A. Tacchella, et al., {\em A  new metrics for Countries Fitness and Products Complexity}, Nature Sci. Rep. \textbf{2}, 723, (2002).

\bibitem{pietronero2} M. Cristelli, et al. {\em Measuring the intangibles: a Metrics for the Economic Complexity of Countries and Products}, PLoS One \textbf{8}, 8, e70726 (2013).

\bibitem{egger} P. Egger, M. Elrich, D. Nelson, {\em Migration and trade}, World Economy \textbf{35}, 216-241, (2012).

\bibitem{head} K. Head, J. Rises, {\em Immigration and Trade creation: Econometric evidence from Canada}, Canad. J. Econ. \textbf{31}:(1), 47--62 (1998).

\bibitem{hunt} J. Hunt, M. Gauthier-Loiselle, {\em How much does immigration boost innovation?}, IZA Disc. PP. $3921$, Institute for the Study of Labor, Bonn (2009).

\bibitem{ozgen} C. Ozgen, P. Nijkamp, J. Poot, {\em Immigration and innovation in European regions}, IZA Disc. PP. 5676 (2011).

\bibitem{partridge} J. Partridge, H. Furtan, {\em Increasing Canada's international competitiveness: is there a link between skilled immigrants and innovation?} Americ. Agricult. Econ. Meeting (Orlando, Florida), (2008).

\bibitem{rashidi} S. Rashidi, A. Pyka, {\em Migration and innovation}, FZID Disc. PP. $77-2013$, (2013).

\bibitem{francisco1} G. Peri, F. Requena-Silvente, {\em The Trade Creation Effect of Immigtants: evidence from the remarkable case of Spain}, Canadian Journ. of Econom.   \textbf{43}(4), 1433-1459, (2010).

\bibitem{francisco2} A. Artal-Tur, V.J. Pallardo'-Lopes, F. Requena-Silvente, {\em The trade-enhancing effect of immigration networks: new evidence on the role of geographics proximity}, Econom. Lett. \textbf{116}, 554-557, (2012).

\bibitem{francisco3} G. Serrano-Domingo, F. Requena-Silvente, {\em Re-examining the migration-trade link using province data: an application of the generalized propensity score}, Econom. Modelling \textbf{32}, 247-261, (2013).

\bibitem{Ghirlanda1} L. Rendell, et al., {\em Why Copy Others? Insights from the Social Learning Strategies Tournament}, Science \textbf{328}, 5975,  208-213 (2010).

\bibitem{Ghirlanda2} A. Bandaura, {\em Social learning theory}, General Learning Press, New York, (1977).

\bibitem{peri} M. Aleksynska, G. Peri, {\em Isolating the Network Effect of Immigrants in Trade},  The World Economy \textbf{37}(3), 434-455, (2014).

\bibitem{gould} D. Gould, {\em Immigrant links to the home country: empirical implications for U.S. bilateral trade flows}, Rev. Econom. $\&$ Statist. \textbf{76}:(2), 302--316 (1994).

\bibitem{caney} T. Chaney, {\em Distorted gravity: The intensive and extensive margins of International Trade}, Amer. Econ. Rev. \textbf{98}:(4), 1707-1721, (2008).

\bibitem{Granovetter-1973} M.S.  Granovetter, {\em The Strength of Weak Ties}, Amer. J. of Sociology \textbf{78},  1360-80, (1973).

\bibitem{Granovetter-1983} M.S. Granovetter, {\em The Strength of the Weak Tie: Revisited}, Sociol. Theory \textbf{1}, 201-33, (1983).

\bibitem{Barra-PhysA2012} A. Barra, E. Agliari {\em A statistical mechanics approach to Granovetter theory}, Physica A, \textbf{391}, 3017--3026 (2012).

\bibitem{Damm-2013}A. P. Damm, Neighborhood Quality and Labor Market Outcomes: Evidence from Quasi-Random Neighborhood Assignment of Immigrants, University Press of Southern Denmark, Odense (2013)

\bibitem{barra0} A. Barra, {\em The mean field Ising model thought interpolating techniques}, J. Stat. Phys. \textbf{132}, 5, 787-809, (2008).

\bibitem{ellis} R.S. Ellis, {\em Large deviations and statistical mechanics}, Springer, New York (1985).

\bibitem{Agliari-EPL2011} E. Agliari, A. Barra {\em A Hebbian approach to complex network generation}, Europhys. Lett. \textbf{4}, 10002, (2011).

\bibitem{Barra-JStat2011} A. Barra, E. Agliari {\em Equilibrium statistical mechanics on correlated random graphs}, J. Stat., P02027 (2011).

\bibitem{bianconi} G. Bianconi, {\em Mean field solution of the Ising model on a Barabasi-Albert network},
Phys. Lett. A. \textbf{303}, 166, (2002).

\bibitem{HerSaa}  M.G. Herander, L. A. Saavedra, {\em Exports and the structure of immigrant-based networks: the role of geographic proximity}, Rev. Econ. Stat. \textbf{87}.2,  323-335, (2005).

\bibitem{pietronero3} G. Caldarelli, et al., {\em A Network Analysis of Countries' Export Flows: Firm Grounds for the Building Blocks of the Economy}, PLoS ONE 7(10): e47278 (2012).

\bibitem{index} T. Beck, et al., {\em New tools in comparative political economy: The Database of Political Institutions},The World Bank Economic Review \textbf{15} (1):165-176, (2001).

\bibitem{marti} A. Barrat, M. Weigt, {\em On the properties of small-world network models}, Europ. Phys. J. B  \textbf{13},  547, (2000).

\bibitem{Agliari-PRE2011} E. Agliari, C. Cioli, E. Guadagnini, {\emph Percolation on correlated random networks}, Phys. Rev. E \textbf{84}, 031120 (2011).

\bibitem{callaway} D. Callaway, M.E.J. Newman, S.H. Strogats, D.J. Watts, {\em Network robustness and fragility: Percolation on random graphs}, Phys. Rev. Lett. \textbf{85}, 5468 (2000).

\bibitem{barabasi} R. Albert, A. L. Barabasi \textit{Statistical mechanics of complex networks}, Rev.  Mod. Phys. \textbf{74}, 47-97 (2002).

\bibitem{vespignani} A. Barrat, M. Barthelemy, A. Vespignani, {\em Dynamical processes in complex networks}, Cambridge University Press (2008).

\bibitem{Estudio} Capital social: confianza, redes y asociacionismo en 13 pais, Unidad de Estudios de Opinion Publica (2006).

\bibitem{HumanOrganization} C. McCarty, P. D. Killworth, H. Russell Bernard, E. C. Johnsen, G. A. Shelley, {\emph Comparing Two Methods for Estimating Network Size}, Human Organization \textbf{60}, 1 (2001).

\bibitem{girma} S. Girma, Z. Yu, {\em The link between Immigration and Trade: Evidence from United Kingdom}, Weltirtschaftliches Archiv \textbf{138}, 1 (2002).



%\bibitem{watts} D.J. Watts, S.H. Strogatz, {\em Collective dynamics of small world networks}, Nature \textbf{393}, 440--442, (1998).

%\bibitem{watts2} P.S. Dodds, R. Muhamad,  D.J. Watts, {\em An experimental study of search in global social networks. science}, \textbf{301}, 827-829, (2003).

%\bibitem{achlioptas} D. Achlioptas, R. M. D'Souza, J. Spenceer, {\em Explosive Percolation in Random Networks}, Science \textbf{323}, 1453 (2009).

%\bibitem{ABC} E. Agliari, A. Barra, F. Camboni, {\em Criticality in diluted ferromagnets}, J. Stat. Mech., (2008).

%\bibitem{Rad} A. Ajadari Rad, et al., {\em Topological measure locating the effective crossover between segregation and integration in a modular network}, arXiv:1206:3403v1

%\bibitem{bollo} B. Bollobas, {\em Random graphs}, Cambridge Stud. Advanc. Math., Cambr. Univ. Press. (1985).

%\bibitem{borias} G.J. Borias, {\em The economic benefits from Immigration}, J. Econ. Perspect. \textbf{9}:(2), 3--22 (1995).

%\bibitem{bouchaud} J.P. Bouchaud, M. Potters, \textit{Theory of financial risks: from statistical physics to risk managemenent}, Oxford University Press, (2004).

%\bibitem{durlauf1} W. Brock, S. Durlauf, {\em Discrete choices with social interactions}, Rev. Econ. St. \textbf{68}, 2, 235-260, (2001).

%\bibitem{bucca} M. Buchanan, \textit{Nexus: Small Worlds and the Groundbreaking Theory of Networks}. Norton, W. W. Company, Inc. (2003).

%\bibitem{castellano} C. Castellano, S. Fortunato, V. Loreto, \emph{Statistical physics of social dynamics}, Rev.\ Mod.\ Phys. 81, 591-646 (2009).

%\bibitem{coleman} D. Coleman, R. Rowthorn, {\em The economic effects of Immigration into the United Kingdom}, Popul. and Developm. Rev. \textbf{30}:(4), 579-624, (2004).

%\bibitem{coppel} J. Coppel,  J.C. Dumont, I. Visco,  {\em Trends in immigration and economic consequences},  ECO WKP-10 (2001).

%\bibitem{durlauf2} S.N. Durlauf, {\em How can statistical mechanics contribute to social science?}, Proc. Natl. Ac. Sc. \textbf{96}, 19, 10582-10584, (1999).

%\bibitem{gao}  J. Gao,  et al., {\em Networks formed from interdependent networks}, Nature Phys. \textbf{8}, 40, (2012).

%\bibitem{hebb} D.O. Hebb, {\em Organization of Behavior}, Wiley, New York, (1949).

%\bibitem{hopfield} J.J. Hopfield, {\em Neural networks and physical systems with emergent collective computational abilities}, P.N.A.S. \textbf{79}, (1982).

%\bibitem{fadden} D. McFadden, {\em Economic choices}, American Econ. Rev. \textbf{91}, 351-378, (2001).

%\bibitem{MPV} M. M\'ezard, G. Parisi and M. A. Virasoro, {\em Spin glass theory and beyond}, World Scientific, Singapore (1987).

%\bibitem{milgram} S. Milgram, {\em The Small World Problem}, Psych. Today \textbf{2}, 60-67, (1967).

%\bibitem{sole} J. M. Montoya, R. V. Sole', {\em Small World Patterns in Food Webs}, J. Theor. Biol. \textbf{214}, 3, (2002).

%\bibitem{ton2} T. Nikoletopoulos, A.C.C. Coolen, I. Perez-Castillo, N.S. Skantzos, J.P.L. Hatchett, B. Wemmenhove, {\em Replicated Transfer Matrix Analysis of Ising Spin
%Models on `Small World' Lattices}, J. Phys. A \textbf{37}, (2004).

%\bibitem{plenz} S. Pajevic,  D. Plenz,  {\em The organization of strong links in complex networks}, Nature Phys. \textbf{8}, 429, (2012).

%\bibitem{Albert-RevModPhys2002} R. Albert, A.L. Barabasi, {\em Statistical mechanics of complex networks}, Rev.\ Modern Phys. \textbf{74}, 47--97  (2002).

%\bibitem{Newman-SIAM2003} M.E.J. Newman, {\em The structure and function of complex networks}, SIAM Rev. \textbf{45}, 167--256 ,(2003).
%and references therein






\end{thebibliography}
\end{document}